\documentclass[pra,twocolumn,groupedaddress]{revtex4-1}
\usepackage{graphicx}
\usepackage{dcolumn}
\usepackage{bm}

\usepackage{amsmath}
\usepackage{graphicx}
\usepackage{color}
\usepackage{wrapfig}
\usepackage{epstopdf}

\graphicspath{{pics/}}

\newcommand{\bee}{\begin{equation}}
\newcommand{\ene}{\end{equation}}

\newcommand{\bea}{\begin{eqnarray}}
\newcommand{\ena}{\end{eqnarray}}

\begin{document}

\title{All-dielectric nanophotonics: fundamentals, fabrication, and applications}

\author{Alexander~Krasnok$^{a,b,*}$, Roman~Savelev$^{b}$, Denis~Baranov$^{c,d}$, and Pavel~Belov$^b$}
\address{$^a$The University of Texas at Austin, Austin, Texas 78712, USA\\
$^b$ITMO University, St.~Petersburg 197101, Russia\\
$^c$Moscow Institute of Physics and Technology, Dolgoprudny 141700, Russia\\
$^d$Chalmers University of Technology, 412 96 Gothenburg, Sweden\\
$^*$E-mail: krasnokfiz@mail.ru}

\begin{abstract}
In this Article, we review a novel, rapidly developing field of modern light science named all-dielectric nanophotonics. This branch of nanophotonics is based on the properties of high-index dielectric nanoparticles which allow for controlling both magnetic and electric responses of a nanostructured matter. Here, we discuss optical properties of high-index dielectric nanoparticles, methods of their fabrication, and recent advances in practical applications, including the quantum source emission engineering, Fano resonances in all-dielectric nanoclusters, surface enhanced spectroscopy and sensing, coupled-resonator optical waveguides, metamaterials and metasurfaces, and nonlinear nanophotonics.
\end{abstract}

\maketitle

\section{Introduction}\label{DN_introduction}

Since the modern technologies largely depend on the rapidly growing demands for powerful computational capacities and efficient information processing, the development of conceptually new approaches and methods is extremely valuable. One of these approached is based on replacing electrons with photons as the main information carriers~\cite{CaulfieldNP2010}. The advantages of light for fast computing are obvious: the parallel transfer and processing of signals using polarization and orbital momentum of photons as additional degrees of freedom~\cite{Willner_2014}, the possibility of multi-frequency operations, and the high operating frequency around 500~THz (wavelength of 600~nm). However, photons as alternative information carriers have relatively large ``size'' determined by their wavelength. This leads to a weak interaction of photons with nanoscale objects including quantum emitters, subwavelength waveguides, and others, whereas the effective light-matter coupling is extremely important for all-optical information processing.

The efficient light manipulation implies simultaneous control of its electric and magnetic components. However, the magnetic response of natural materials at optical frequencies is usually weak, as was originally posted by Landau and Lifshitz~\cite{Landau_CM, Merlin}. This is the reason why photonic devices operate mainly with the electric part of a light wave~\cite{Kuipers}. At the same time, magnetic dipoles are very common sources of the magnetic field in nature. A common example of a magnetic dipole radiation is an electromagnetic wave produced by an excited metal split-ring resonator (SRR), which is a basic constituting element of \textit{metamaterials}~\cite{Soukoulis, Pendry, Shelby}. Currents excited by the external electromagnetic radiation and running inside the SRR produce a transverse magnetic field in the center of the ring oscillating up and down, which simulates an oscillating magnetic dipole. The major interest in such artificial systems is due to their ability to response to the magnetic component of incident light and thus to have a non-unity or even negative magnetic permeability ($\mu$) at optical frequencies. This provides the possibilities to design materials with highly unsusual properties such as negative refraction~\cite{Shalaev, Zheludev, Soukoulis, KivsharNM2012}, cloaking~\cite{Leonhardt}, or superlensing~\cite{Pendry00}. The SRR concept works very well for gigahertz~\cite{Smith, Shelby}, terahertz~\cite{Padilla06} and even near-infrared~\cite{Liu08} frequencies. However, this approach fails for shorter wavelengths and, in particular, in the visible spectral range due to increasing losses and technological difficulties in fabrication of smaller and smaller constituting split-ring elements~\cite{Soukoulis07}. Several other designs based on metal nanostructures have been proposed to shift the magnetic resonance wavelength to the visible spectral range~\cite{Shalaev, Zheludev}. However, all of them are suffering from losses inherent to metals at visible frequencies.

\begin{figure}[!b] \centering
\includegraphics[width=0.99\columnwidth]{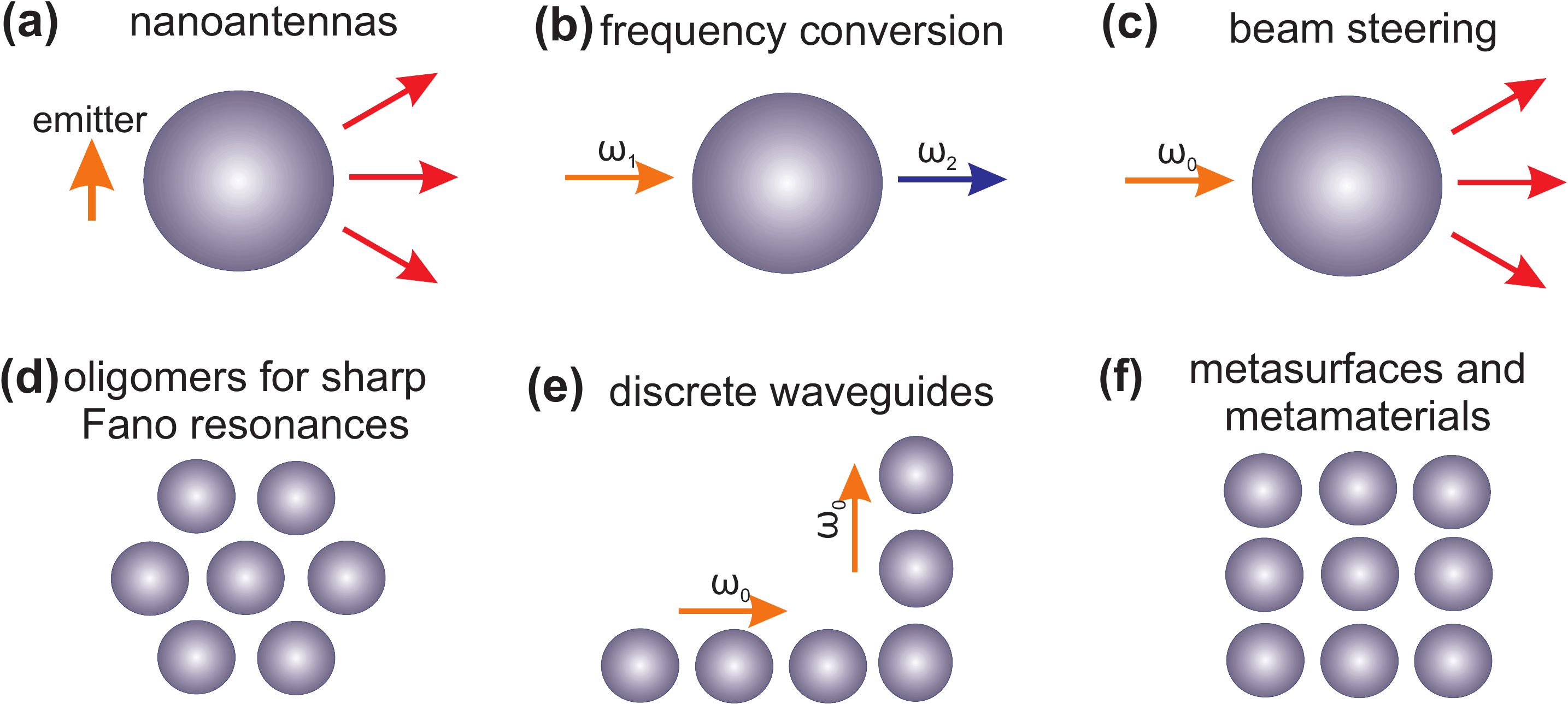}
\caption{The variety of photonics structures based on high-indexed dielectric nanoparticles are discussed in this Review: (a)~ Nanoantennas for enhanced emission and focusing; (b)~Nanoantennas for enhanced frequency conversion effects; (c)~Nanoantennas for steering of light; (d)~Fano-resonant nanostructures; (e)~Waveguides composed of dielectric nanoparticles; (f)~All-dielectric metasurfaces and metamaterials.}
\label{fig:alldielnano}
\end{figure}

An alternative approach to achieve strong magnetic response with low losses is to use nanoparticles made of high-refractive index dielectric materials (e.g. Si, Ge)~\cite{Peng2007, ShullerPRL2007, Meng2008, Guizal2009, Zhao09, Cummer_08, Brener_12, Lukyanchuk13, StaudeACSNano2013, Zhou2014, Habteyes2014, Savelev2015, Baranov16, Dmitriev_2016}. As it follows from the exact Mie theory of light scattering~\cite{Bohren}, a strong magnetic dipole response can be achieved in spherical dielectric particles. Remarkably, for the refractive indices larger than a certain value, there is a well-established hierarchy of the magnetic and electric resonances. In contrast to plasmonic particles, the fundamental resonance of high-index spherical nanoparticles is the \textit{magnetic dipole} one, occurring when the wavelength of light inside the particle approximately equals its diameter $\lambda/n\simeq2 R$, where $\lambda$ is the wavelength in free space, $R$ and $n$ are the radius and the refractive index of spherical particle. Under this condition, the polarization of the electric field is anti-parallel at opposite boundaries of the sphere, which gives rise to strong coupling to circulation displacement currents, while the magnetic field oscillates up and down in the middle. Today, such high-index dielectric nanoparticles are the base of so-called \textit{all-dielectric nanophotonics}.

It should be noted that the all-dielectric nanophotonics is not a part of so-called \textit{silicon photonics}~\cite{Soref2006, Daldosso2009} since it does not limited by silicon as a material. Moreover, the silicon photonics is based on silicon waveguides and circuits with dimensions of few hundred nanometers, whereas the heart of all-dielectric nanophotonics is high-index nanoparticles exhibiting a strong magnetic optical response.

The prediction of the resonant electromagnetic response of small dielectric particles is not an absolutely new discovery. These resonances have been recognized since the original work of G.~Mie~\cite{Mie}. Later, in 1930 J.~A.~Stratton has highlighted the effect of magnetic and electric dipole resonances of water drops in the atmosphere (rain, fog, or clouds) on the propagation of short radio waves~\cite{Stratton}. In the following studies (see, e.g., Ref.~\cite{Lewin}) the theoretical prediction of man-made artificial dielectrics based on high-index particles was made. Then, the metamaterials based on the dielectric particles have been proposed in the microwave and mid-IR frequency ranges~\cite{Peng2007, ShullerPRL2007, Cummer_08}.

In 2010 the possibility of enhanced optical magnetic response of high-index nanoparticles in the visible range was discussed theoretically in Ref.~\cite{EvlyukhinPRB2010}. It should be noted that before this work, a number of papers have been published, where the spectra of light scattering by high-index dielectric cylinders including magnetic resonances were measured~\cite{Schuller2009, Cao2010, Brongersma_NL_10, Brongersma_10}. However, an attention to the magnetic nature of these resonances has not been paid. In 2011 the scattering properties of silicon (Si) nanoparticles have been studied in details~\cite{Garcia-Etxarri2011}. Shortly after these theoretical works, the concept of "magnetic light" has been experimentally realized in visible~\cite{KuznetsovSciRep2012, Vesseur2012, EvlyukhinNL2012}, infrared~\cite{ShiAdvMat2012}, and even microwave~\cite{Geffrin:NC:2012} frequency ranges.

This magnetic light concept has paved the way for many fascinating applications in the areas of quantum source emission engineering~\cite{KrasnokOE2012, Miroshnichenko:NL:2012, Chong2014, Filonov_oligomer, KrasnokNanoscale, Krasnok_LPR_2015, Li2015, Bogdanov_2016, Markovich2016} (Figure~\ref{fig:alldielnano}(a)), frequency conversion~\cite{ShcherbakovNL2014} (Figure~\ref{fig:alldielnano}(b)), tunable routers and switchers~\cite{Noskov2012, Baranov_2016_ACS, Baranov_dimer_2016} (Figure~\ref{fig:alldielnano}(c)), sensors~\cite{Granzow2013, Guo2014, Huang2015} (Figure~\ref{fig:alldielnano}(d)), dielectric waveguides~\cite{Savelev2014_1, Savelev_PRB_2015, Savelev_ACSP_2016} (Figure~\ref{fig:alldielnano}(e)), and all-dielectric metasurfaces~\cite{Staude_15, Tsypkin_2016}, and metamaterials~\cite{Cummer_08, Zhao09, Brener_12, ShiAdvMat2012, Valentine2014} (Figure~\ref{fig:alldielnano}(f)). All these applications are discussed in this Review (see Sec.~\ref{sec:Applications}).
\begin{figure}[!t] \centering
\includegraphics[width=0.99\columnwidth]{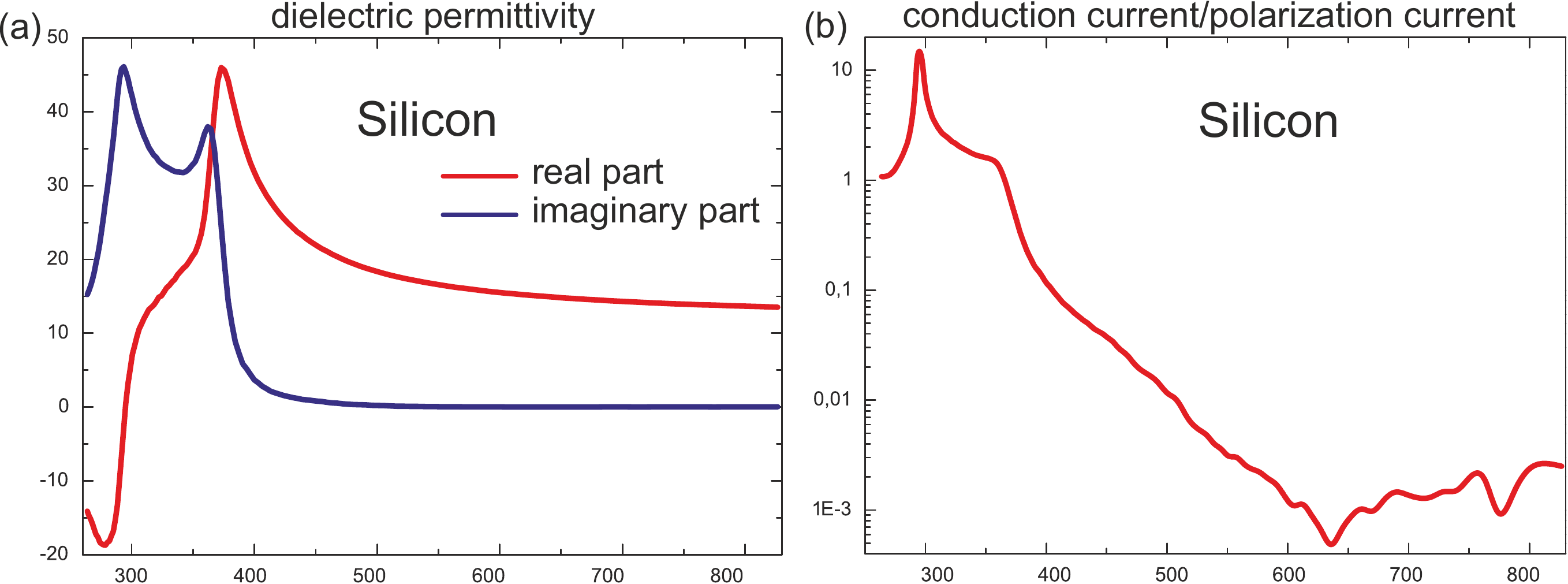}
\caption{(a)~Real and imaginary parts of the permittivity of crystalline silicon~\cite{VuyeSi}. (b)~Ratio of the conductivity and displacement currents in crystalline silicon. The displacement current strongly exceeds the conductivity current at wavelengths above 400~nm -- in this spectral range the pure and crystalline Si can be considered as dielectric.}
\label{fig:epsSi}
\end{figure}

This Article is intended for review the all-dielectric nanophotonics as a part of the modern light-science. First, in Sec.~\ref{sec:OpticalProperties} we discuss the optical properties of high-index dielectric nanoparticles with an example of silicon nanoparticles. Then, in Sec.~\ref{sec:fabrication} we describe the various existing methods of such nanoparticles fabrication, including the chemical deposition, the thin film dewetting, the femtosecond laser ablation of bulk and thin Si films, and the reactive-ion-etching approach. The examples of the produced structures are presented. In Sec.~\ref{sec:nanoantennas} the applications of all-dielectric nanoantennas for the quantum source emission engineering are presented. In Sec.~\ref{sec:fano} the all-dielectric oligomers and their Fano resonances are discussed. Then in Sec.~\ref{sec:sensing} the applications of dielectric nanoatennas for the surface enhanced spectroscopy and sensing are provided. The all-dielectric nanoantennas and plasmonic ones are compared in the context of the sensing applications. In Sec.~\ref{sec:waveguides} the properties of coupled-resonator nanoparticle waveguides are discussed. The optical solitons and bound-states-in-continuum in the dielectric waveguides are also reviewed. Sec.~\ref{sec:metamaterials} is devoted to the brief review of the all-dielectric metamaterials and metasurfaces. The references on the more comprehensive reviews of this topic are proposed. In Sec.~\ref{sec:nonlinear} the nonlinear properties of silicon nanoparticles with electric and magnetic dipole responses, including higher harmonics generation and electron-hole plasma photoexcitation, are discussed. In Conclusions the main statements of the Review are summarized and the outlook of this area is given.

\section{Optical properties of high-index dielectric nanoparticles}
\label{sec:OpticalProperties}
\begin{figure*}[t] \centering
\includegraphics[width=0.7\textwidth]{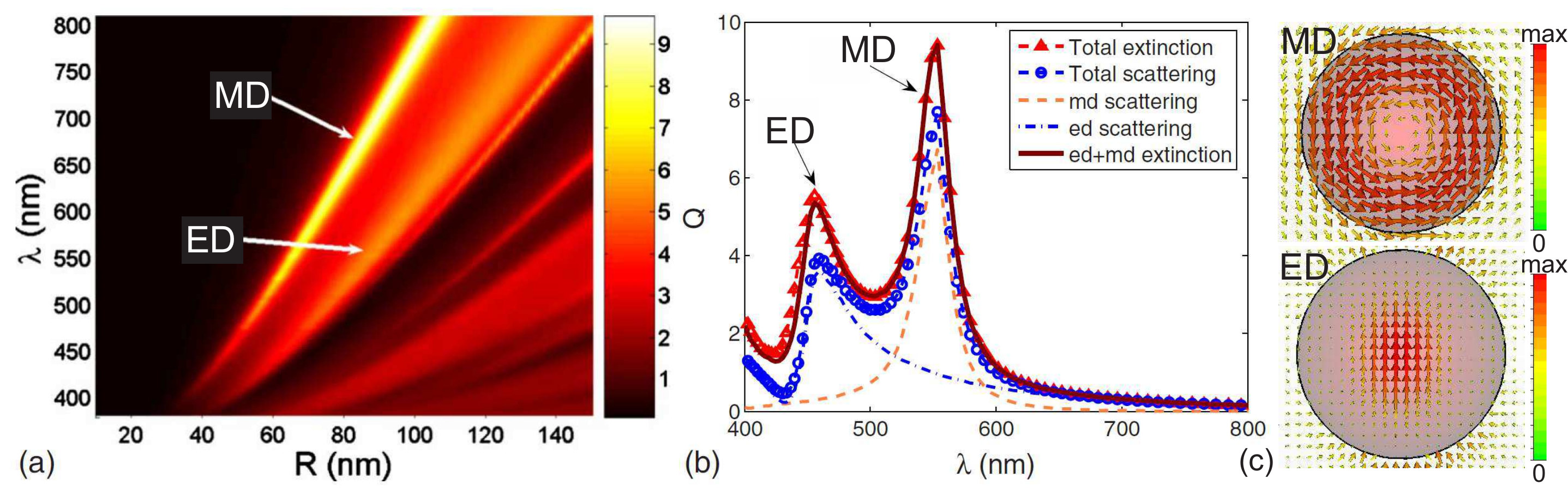}
\caption{(a)~Scattering efficiency spectra of Si spherical particles with the radius $R$ located in air. (b)~Extinction and scattering spectra of a Si particle ($R=65$~nm). The arrows indicate the electric dipole (ED) and magnetic dipole (MD) contributions to the total efficiencies. From the Ref.~\cite{EvlyukhinPRB2010}. (c)~Distributions of the electric field within the nanoparticle at magnetic dipole (MD) and electric dipole (ED) resonances.}
\label{fig:SiNP_extinction}
\end{figure*}

Two types of electric currents appear as the sources of electromagnetic fields in Maxwell equations: the conductivity current and the displacement current~\cite{Novotny_Hecht_book}. In dielectrics and semiconductors, the displacement currents strongly exceed the conductivity currents in the spectral range far from their main absorption band. As an example, the spectral dependencies of real and imaginary parts of the refractive index, measured at room temperatures, are shown in Figure~\ref{fig:epsSi}(a)~\cite{VuyeSi}. In Figure~\ref{fig:epsSi}(b) the ratio of the conductivity and the displacement current is shown. One can see that in silicon the displacement currents strongly exceed the conductivity currents at wavelengths larger 400~nm. Therefore, from the electromagnetic point of view, the pure and crystalline silicon can be considered as a dielectric in the visible spectral range.

A dielectric nanoparticle can be treated as an open resonator supportng a series of electromagnetic resonances -- eigenmodes. Exact analysis of the plane wave diffraction by a spherical particle (Mie scattering) shows that the nanoparticle supports the electric and magnetic eigenmodes of different orders~\cite{Bohren}. The coupling strengths between the incident wave and eigenmodes depend on the size parameter $x=k_{0} n R$, where $n$ is the refractive index of the particle, $k_0$ is the free space wavenumber of the incident radiation, and $R$ is the radius of the particle. If $x \ll 1$, the particle is optically small and its diffraction can be described by the Rayleigh approximation. With increasing $x$, the fundamental magnetic dipole (MD) resonance appears in the particle response. The electric field lines at this resonance are shown in Figure~\ref{fig:SiNP_extinction}(c). The scattered field by the particle at the MD resonance corresponds to the radiation field of a magnetic dipole. With further increase of $x$, the first electric dipole (ED) resonance is formed. For even larger values of the size parameter, the higher order (quadrupole, octupole, etc. moments) multipole modes are excited.

The frequencies of the Mie resonances for a spherical particle in the dipole approximation can be determined from the following conditions:
\begin{eqnarray} \label{eq_Mie_polarizabilities}
\mathrm{Re}(\alpha_e^{-1})&=&\mathrm{Re}[(i\dfrac{3\varepsilon_h}{2k_h^3}a_1)^{-1}]=0,\nonumber\\
\mathrm{Re}(\alpha_m^{-1})&=&\mathrm{Re}[(i\dfrac{3}{2k_h^3}b_1)^{-1}]= 0,
\end{eqnarray}
where $\alpha_e$ and $\alpha_m$ are electric and magnetic polarizabilities, respectively, $\varepsilon_h $ is the host permittivity, $k_h=\sqrt{\varepsilon_h}\omega/c$ is the wavenumber of light in host media, $\omega$ is the angular frequency, $c$ is the speed of light in vacuum, $a_1$ and $b_1$ are the scattering Mie coefficients~\cite{Bohren}. In Ref.~\cite{EvlyukhinPRB2010} it was shown, that for a Si spherical nanoparticles, the conditions of the lowest order multipole (dipole) resonances are fulfilled for the radius of $\approx70$~nm. Figure~\ref{fig:SiNP_extinction} shows the numerically calculated scattering efficiency spectra (a), as well as the extinction and scattering spectra (b) of such particle. Note that the resonance frequencies of the particle can be shifted not only by changing its size, by also its shape~\cite{EvlyukhinPRB2011, EvlyukhinSciRep2014}.

Almost complete absence of conductivity currents in silicon in the optical frequency range leads to low dissipative losses, in contrast to plasmonic structures where the strong field localization is always accompanied by high dissipation. Therefore, by exploiting dielectric particles with the magnetic response one can design different low-loss nanostructures, composite materials and metasurfaces with unique functionalities.

\section{Methods of high-index dielectric nanoparticles fabrication}
\label{sec:fabrication}

Silicon is the most frequently used high-index dielectric in optical and IR ranges owing to its relatively low cost and low imaginary part of the refractive index. Moreover, the technology of fabrication of Si nanoparticles with Mie-resonances has been developing intensively during the last several years, resulting in the emerging of various techniques. The proposed methods of Si nano- and microparticles fabrication can be classified on the level of the particles size and location controllability. Here, we describe the various existing methods of the high-index nanoparticles fabrication, including the \textit{chemical deposition}, the \textit{thin film dewetting}, the \textit{femtosecond laser ablation} of bulk and thin Si films, and the \textit{reactive-ion-etching} approach.

\begin{figure}[!t]\centering
\includegraphics[width=0.99\columnwidth]{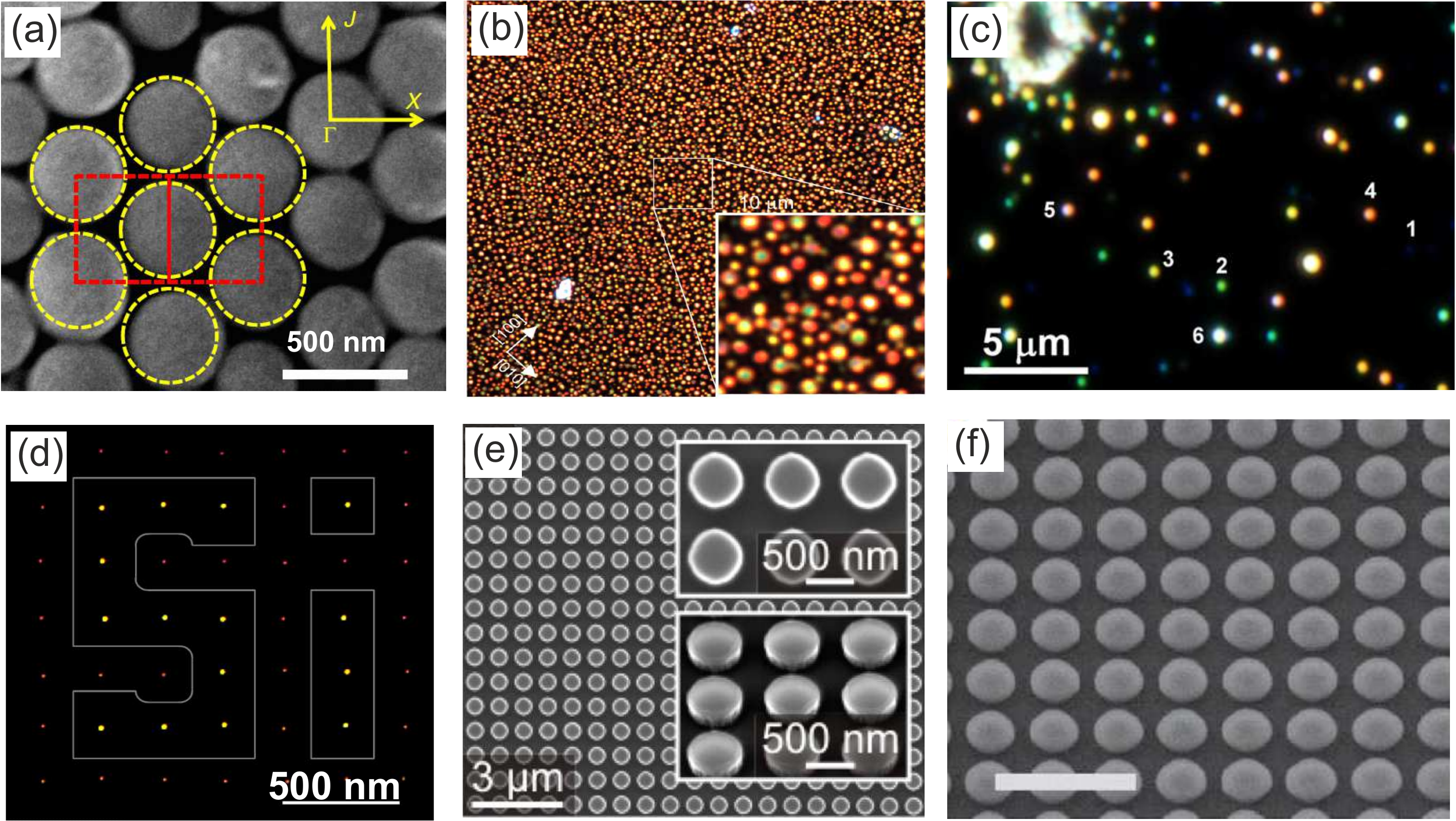}
\caption{(a)~Electron microscopy image of self-aligned Si nanoparticles obtained by chemical deposition~\cite{ShiNC2013}. (b)~Dark-field optical image of Si nanoparticles obtained by thin film dewetting~\cite{AbbarchiACSNano2014}. (c)~Dark-field optical image of Si nanoparticles obtained via femtosecond laser ablation of bulk Si~\cite{KuznetsovSciRep2012}. (d)~Dark-field optical image of Si nanoparticles obtained via femtosecond laser ablation of thin Si film~\cite{Evlyukhin_NC_2014}. In the picture (d)~red nanoparticles are amorphized, while yellow are annealed and crystalline. (e)~Electron microscopy image of Si nanoparticles obtained by means of reactive-ion-etching through a mask~\cite{StaudeACSNano2013}; (f)~the same but with additionally deposited Si$_3$N$_4$ thin film~\cite{SpinelliNC2012}.}
\label{fig:Fabr_1}
\end{figure}

\textbf{Chemical deposition.} The fabrication method of Si nanoparticles with different sizes can be carried out by means of chemical vapor deposition technique, in which disilane gas (Si$_2$H$_6$) decomposes at high temperatures into solid Si and hydrogen gas by the following chemical reaction: Si$_2$H$_6 \rightarrow$ 2Si(s) + 3H$_2$(g). Spherical poly-crystalline Si nanoparticles were produced by this method in Ref.~\cite{ShiAdvMat2012}. Further, fabrication of monodispersed Si colloid was achieved via decomposition of trisilane (Si$_3$H$_8$) in supercritical n-hexane at high temperature in Ref.~\cite{ShiNC2013}. In this advanced method, the particles size can be controlled by changing of trisilane concentration and temperature of the reaction. This relatively simple method allows one to obtain plenty of similar Si nanoparticles with the size dispersion of several percents, which can be ordered into hexagonal lattice via a self-assembly process [Figure~\ref{fig:Fabr_1}(b)]. The main disadvantage of this method is the porosity and high hydrogen content in each nanoparticle as well as the necessity of their additional ordering to fabricate functional structures.

\textbf{Thin film dewetting.} Disordered Si nanoparticles of different sizes can be also produced via dewetting of thin Si film after its heating [Figure~\ref{fig:Fabr_1}(b)]~\cite{AbbarchiACSNano2014}. In this case, the nanoparticles can be crystalline and their sides are aligned along crystallographic facets. The main controlling parameters in this method are the heating temperature and the film conditions (defects and initial pattern)~\cite{AbbarchiACSNano2014}. In thin film dewetting technique, the control over the nanoparticles size and location can be achieved only by using additional lithographical methods, which is even more complicated in comparison with the chemical deposition techniques. Indeed, the chemical deposition and thin film dewetting methods are more suitable for the high-throughput and low-cost nanoparticles fabrication.

\textbf{Femtosecond laser ablation.} In order to improve the control over the location of fabricated nanoparticles, the laser ablation by focused beam can be used. Indeed, an ultrashort laser pulse focused on the Si surface can heat the material up to the critical point, leading to the material fragmentation into spherical nanoparticles and their deposition nearby the heated area~\cite{KuznetsovSciRep2012, Lukyanchuk13} [Figure~\ref{fig:Fabr_1}(c)]. It worth noting that the colloids of chemically pure nanoparticles can be obtained by means of the laser ablation as well as the chemical deposition. The main advantages of the ablation approach are the high-productivity and the lack of harmful chemical waste.

Fabrication of Si nanoparticles, demonstrating Mie-resonances in the visible range, with accurate control over their positions was carried out by  focusing the femtosecond laser onto the Si surface, from which  nanoparticles were emitted to the transparent receiver substrate~\cite{EvlyukhinNL2012, ZywietzAPA2014, Evlyukhin_NC_2014} [Figure~\ref{fig:Fabr_1}(d)]. There are three main parameters that affect ablated Si nanoparticles: laser intensity, beam spatial distribution and sample thickness. For instance, a single Si nanoparticle with a certain size can be formed from bulk Si after irradiation by a single laser pulse with ring-type spatial distribution~\cite{ZywietzAPA2014}, or after irradiation of thin Si film by conventional Gaussian beam~\cite{Evlyukhin_NC_2014}. The ultrashort laser can be used not only for fabrication but also for Si nanoparticles postprocessing. In particular, the well-known effect of the laser annealing was applied for Si nanoparticles in order to controllably change them from the initially amorphized state to crystalline one, thus tailoring their optical properties [Figure~\ref{fig:Fabr_1}(d)]~\cite{Evlyukhin_NC_2014}.

\textbf{Reactive-ion-etching.} The most controllable fabrication of Si nanoparticles was achieved by a multi-stage method, including electron-beam lithography on Si-on-insulator wafers (formation of a mask from resist) and the reactive-ion-etching process with following removing of the remaining electron-beam resist mask. This advanced technology enables the formation of Si nanocylinders [Figure~\ref{fig:Fabr_1}(e)], in which Mie-resonances can be precisely tuned by varying the basic geometrical parameters (diameter and height). Various types of structures based on the Si nanocylinders have been designed in order to show unique properties of the all-dielectric nanophotonics devices~\cite{SpinelliNC2012, PersonNL2013, StaudeACSNano2013, ShcherbakovNL2014}. To achieve higher absorption of the fabricated Si metasurface, this method was supplemented by a deposition of Si$_3$N$_4$ thin film~\cite{SpinelliNC2012} [Figure~\ref{fig:Fabr_1}(f)]. Note that lithography-based methods have such serious disadvantages as high-cost and low-productivity of technological process in comparison with above mentioned lithography-free methods.

In order to summarize the section on methods of high-index dielectric nanoparticles fabrication, we would like to emphasize that currently developing approaches allow one to create various types of all-dielectric functional structures with given optical properties, which will be discussed below.

\section{All-dielectric nanophotonics applications}
\label{sec:Applications}

\subsection{Quantum source emission engineering}
\label{sec:nanoantennas}

The recently emerged field of optical nanoantennas is promising for its potential applications in various areas of nanotechnology. The ability to effectively emit light in a desired direction, redirect propagating radiation, and transfer it into localized subwavelength modes at the nanoscale~\cite{Novotny_10_NatPhot} makes \textit{optical nanoantennas} highly desirable for many applications. Originally, antennas were suggested as sources of electromagnetic radiation at radio frequencies and microwaves, emitting radiation via oscillating currents. Different types of antennas were suggested and demonstrated for the effective manipulation of the electromagnetic radiation~\cite{Balanis}. Thus, conventional antennas perform a twofold function as a source and transformation of electromagnetic radiation, resulting in their sizes being comparable with the operational wavelength. The recent progress in the fabrication of nanoscale elements allows bringing the concept of the radio frequency antennas to optics, leading to the development of optical nanoantennas consisting of subwavelength elements~\cite{Novotny_10_NatPhot}. Currently, nanoantennas are used mainly for near-field microscopy~\cite{Fan:NL:2012}, high resolution biomedical sensors~\cite{Zhang:NL:2011}, photovoltaics~\cite{Spinelli2012}, and medicine~\cite{rac_10}. This section is devoted to the review of nanoantennas based on all-dielectric nanoparticles.

\begin{figure*}[!t]\centering
\includegraphics[width=0.99\textwidth]{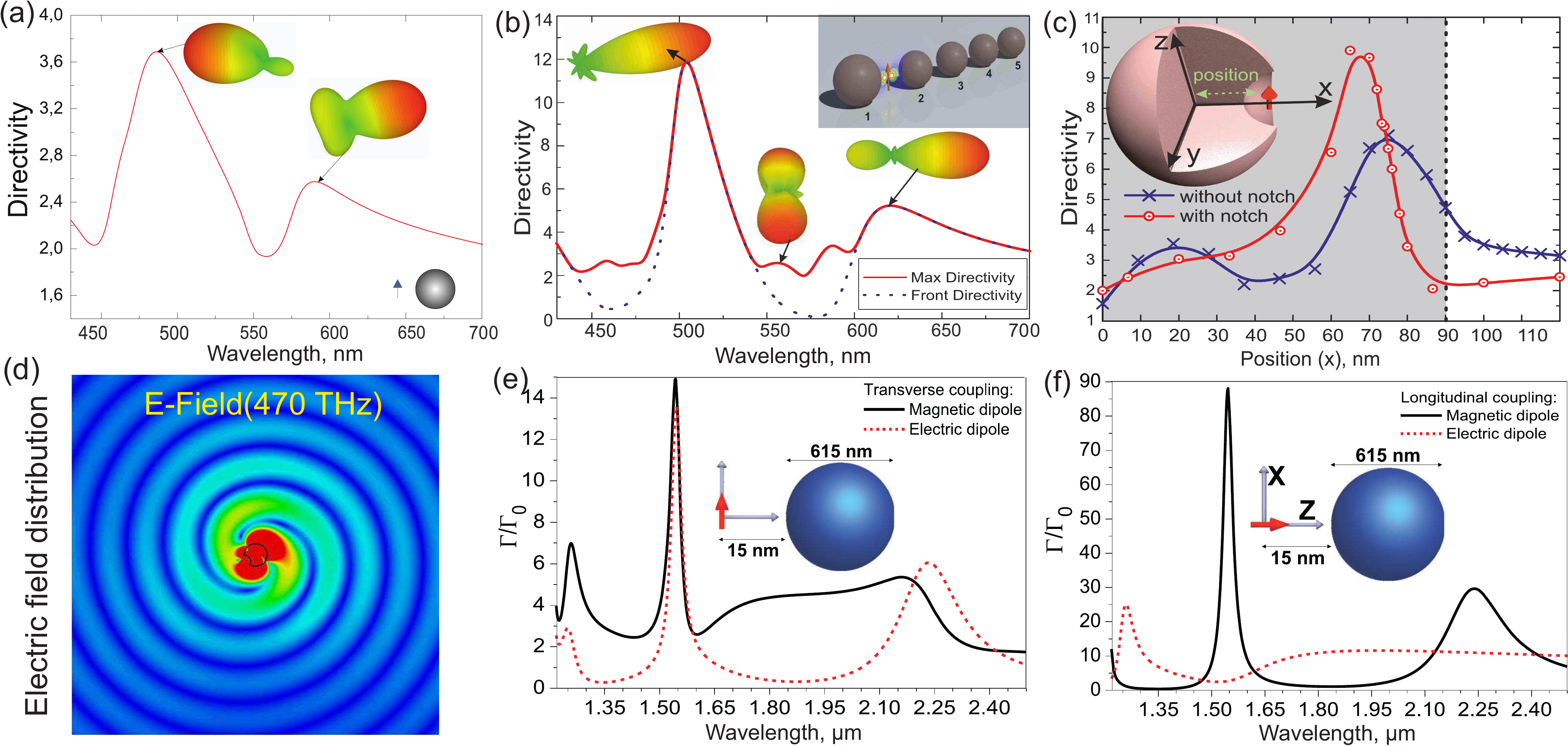}
\caption{(a)~Directivity of the all-dielectric nanoantenna \textit{Huygens element}, consisting of a Si nanoparticle and poind dipole source vs. the radiation wavelength~\cite{Krasnok2011, Krasnok2012}. (b)~Directivity of the all-dielectric \textit{Yagi-Uda nanoantenna}, consisting of the reflector with the radius of $R_r=75$~nm, and smaller directors with the radii of $R_d=70$~nm vs. the radiation wavelength~\cite{Krasnok2012}. Insert: 3D radiation patterns at particular wavelengths. (c)~Maximum of the directivity of the \textit{superdirective nanoantenna} depending on the position of the dipole source at the wavelength of 455 nm in the case of the Si nanosphere with (red curve) and without (blue curve) notch~\cite{Krasnok2014}. Inset: the geometry of the superdirective nanoantenna and the dipole source location in the notch. (d)~Near field of the all-dielectric nanoantenna for chiral near-field formation and unidirectional excitation of electromagnetic guided modes~\cite{Li2015}. (e) and (f)~Normalized decay rates ($\Gamma/\Gamma_{0}$) as a function of the wavelength, for an emitter placed at 15~nm from the surface of a Si sphere of diameter 615~nm for transverse and longitudinal orientation of the dipole, respectively; the surrounding medium is air~\cite{Rolly2012}.}
\label{fig1}
\end{figure*}

Apparently, the first work on the all-dielectric nanoantennas was Ref.~\cite{Schuller_09}. However, in this article, specific implementations of such nanoantennas have not been proposed. The first layout of \textit{all-dielectric nanoantennas} has been proposed in Ref.~\cite{Krasnok2011} in 2011. In this work it has been shown that one Si nanoparticle can have the properties of a \textit{Huygens element} in the optical wavelength range (Figure~\ref{fig1}(a)). It has been demonstrated, that such nanoantennas are able to switch the radiation pattern between forward and backward directions due to the presence of the electric and magnetic resonant modes. Then, in the Ref.~\cite{Krasnok2012} it has been shown that Si nanoparticles can be arranged in the \textit{Yagi-Uda geometry} for creating highly efficient and directive optical nanoantennas (see Figure~\ref{fig1}(b)). The radii of the directors and the reflector have been chosen to achieve the maximal constructive interference in the forward direction along the directors array. The optimal performance of the Yagi-Uda nanoantenna has been achieved for the radii of the directors correspond to the magnetic resonance ($R_d=70$~nm), and the radius of the reflector correspond to the electric resonance ($R_r=75$~nm) at a given frequency. In Figure~\ref{fig1}(b) the directivity of the all-dielectric Yagi-Uda nanoantenna as a function of wavelength for the separation distance between the particles of $D=70$~nm is presented. Inserts demonstrate the 3D radiation patterns at particular wavelengths. A strong maximum of the nanoantenna directivity at $\lambda = 500$~nm has been achieved. In this resonant regime the nanoantenna has the directivity about 12 and the extremely narrow main lobe about $40^\circ$. The maximum of the dependence does not correspond exactly to either magnetic or electric resonances of a single dielectric sphere, which implies the importance of the interaction between constitutive nanoparticles. By comparing plasmonic and the all-dielectric Yagi-Uda nanoantennas, it has been demonstrated that the all-dielectric ones may exhibit better radiation efficiency also allowing more compact design.

These ideas were developed in the series of subsequent works~\cite{Rolly2012, Krasnok_APL_12, RollyOE12, Yang12OL, Schmidt2012, Maksymov2012, Zou2013, Liu2014, Krasnok_SR2015}. It has been demonstrated that the unique optical properties and low dissipative losses make dielectric nanoparticles perfect candidates for design of high-performance nanoantennas, low-loss metamaterials, and other novel all-dielectric nanophotonic devices. The key to such novel functionalities of high-index dielectric nanophotonic elements is the ability of subwavelength dielectric nanoparticles to support simultaneously both electric and magnetic resonances, which can be controlled independently for particles of non-spherical forms~\cite{StaudeACSNano2013}. Moreover, the magnetic field localization in a vicinity of all-dielectric nanoantennas has been theoretically proposed~\cite{Albella2013} and experimentally realized in the microwave~\cite{Boudarham2013} and in the visible~\cite{Bakker2015} ranges.

Generally speaking, achieving of a high radiation directivity is a very important issue for the field of optical nanoantennas~\cite{Novotny_10_NatPhot}. One of the most straightforward ways to achieve high directivity is to combine the nanoantenna constituent elements in the described above Yagi-Uda geometry. However, there is one more general approach consisting in the excitation of high-order multipole modes in the nanoantenna. In Refs.~\cite{KrasnokNanoscale, KrasnokAPLSuperdir} the novel concept of the so-called \textit{superdirective nanoantennas} based on the excitation of higher-order magnetic multipole moments in subwavelength dielectric nanoparticle has been proposed (see Figure~\ref{fig1}(c)). The superdirective regime has been achieved by placing an emitter (e.g. a quantum dot) within a small notch created on the nanosphere surface. The notch has the shape of a hemisphere. The emitter is shown in Figure~\ref{fig1}(c) by a red arrow. It turns out that such a small modification of the sphere allows for the efficient excitation of higher-order spherical multipole modes. Figure~\ref{fig1}(c)~shows the dependence of the directivity maximum $D_{\rm max}$ on the position of the emitting dipole in the case of a sphere $R_{\rm s}=90$~nm without a notch, at the wavelength $\lambda=455$~nm (blue curve with crosses). This dependence has the maximum ($D_{\rm max}=7.1$) when the emitter is placed inside the particle at the distance 20~nm from its surface. The multipole decomposition analysis has shown that, in this case, the electric field distribution inside a particle corresponds to the noticeable excitation of higher-order multipole modes. This becomes possible due to a strong inhomogeneity of the external field produced by the emitter. Furthermore, it has been demonstrated that the excitation of higher-order multipoles can be significantly improved by making a small notch on the Si nanoparticle surface and placing the emitter inside that notch, as shown in Figure~\ref{fig1}(c). The notch has the form of a hemisphere with the center at the dielectric nanoparticle surface. The optimal radius of the notch $R_{\rm n}=40$~nm has been found by means of numerical optimization. The red curve with circles in Figure~\ref{fig1}(c) shows the directivity maximum dependence on the dipole source location for this geometry. The maximal directivity at the wavelength of 455~nm is $D_{\rm max}=10$. Thus, Figure~\ref{fig1}(c) demonstrates the importance of the notch presence in order to achieving the higher directivity of the source radiation. Note that the proposed superdirectivity effect is not associated with high dissipative losses, because of the generally magnetic nature of the nanoantenna operation.

Recently, study of nanoantennas for formation of chiral distributions of the near-field has gained considerable interest~\cite{Huang2014, Zhan2010, Giessen12}. In particular, in Ref.~\cite{Giessen12} it was shown that the chiral near-field can be produced by a symmetric non-chiral nanoantenna. In work~\cite{Huang2014} the chiral distribution of the near-field was investigated in the context of trapping and rotation of nanoparticles. In paper~\cite{Chan2014} it was demonstrated that excitation of the chiral near-field leads to the emergence of lateral optomechanical force acting on a chiral particle. Moreover, such nanostructures enable the generation of light beams with orbital angular momentum~\cite{Capasso14NM}. These effects are the consequence of the fundamental coupling between the spin angular momentum of an evanescent wave and the direction of its propagation, known in the literature as the \textit{spin-orbit coupling}~\cite{Bliokh2012, Bliokh2014, VanMechelen2016}. In Ref.~\cite{Li2015}, the asymmetric excitation of high-index dielectric subwavelength nanoantenna by a point source, located in the notch at the nanoantenna surface has been studied. The generation of the chiral near-field similar to that of a circularly polarized dipole or quadrupole depending on the frequency of the driving source has been demonstrated (see Figure~\ref{fig1}(d)). Using numerical simulations, it has been shown that this effect is the result of the higher multipole modes excitation within the nanoantenna. In this work it has been demonstrated that this effect can be applied for the unidirectional launching of waveguide modes in the dielectric and plasmonic waveguides. Contrary to the strategy employed in Refs.~\cite{Ginzburg_13, Capasso2013PRB}, the directional launching of the guided modes achieved without a rotating or a circularly polarized point dipole source, but due to the violation of the rotational symmetry of the system.

One more important feature of the optical nanoantennas is their ability to exhibit strong \textit{Purcell effect}. The Purcell effect is manifested in a modification of the spontaneous emission rate ($\Gamma$) of a quantum emitter induced by its interaction with inhomogeneous environment and is quantitatively expressed by the Purcell factor~\cite{Purcell_46, Sauvan2013a, BarthesPhysRevB2011, Poddubny2012, Krasnok_Purcell_2015, KrasnokAPL2016}. This modification is significant if the environment is a resonator tuned to the emission frequency. Open nanoscale resonators such as plasmonic nanoantennas can change the spontaneous emission lifetime of a single quantum emitter, that is very useful in microscopy of single NV centers in nanodiamonds~\cite{Vamivakas_NanoLetters_13}, Eu$^{3+}$-doped nanocrystals~\cite{Carminati_PRL_14}, plasmon-enhanced optical sensing~\cite{Sauvan2013a}, and the visualization of biological processes with large molecules~\cite{Tinnefeld_Science_2012}.

Mie resonances in dielectric particles can also increase the Purcell factor associated with either electric or magnetic transition rates in nearby quantum emitters. Their large quality factors compensate their low field confinement as compared to the plasmon resonances of metallic nanostructures for which nonradiative decay channels dominate. In Ref.~\cite{Rolly2012} it has been shown theoretically that near-infrared quadrupolar magnetic resonances in Si nanoparticles can preferentially promote magnetic versus electric radiative deexcitation in trivalent erbium ions at 1.54~$\mu$m (see Figure~\ref{fig1}(e,f)). The distance-dependent interaction between magnetic (electric) dipole emitters and induced magnetic or electric dipoles and quadrupoles has been derived analytically and compared to quasiexact full-field calculations based on Mie theory. The detailed analysis of the Purcell effect in the plasmonic and all-dielectric nanoantennas is presented in Sec.~\ref{sec:sensing}.

\subsection{Fano resonances in all-dielectric oligomers}
\label{sec:fano}

The Fano resonance~\cite{Fano_61,Miroshnichenko:NL:2012,Liu2015} is known to originate from the interference of two scattering channels, one of which is non-resonant, while the other is strongly resonant. Fano resonance was observed in different areas of physics, including photonics, plasmonics, and metamaterials~\cite{Lukyanchuk:NM:2010}. It has been demonstrated that the Fano resonance is highly sensitive to the optical properties of the background medium, which makes it perspective for the design of sensors. In the last few years, there is a growing interest in studying the Fano resonances in the so-called plasmonic \textit{oligomer structures}, that consist of several symmetrically positioned metallic nanoparticles~\cite{Rahmani:NL:2012, Zhang:PNAS:2013}. In such structures, the Fano resonance appears as a resonant suppression of the scattering cross-section of the structure, and it is accompanied by a strong near-field enhancement and, consequently, absorption.

Recently, it was shown that the oligomers composed of high-index dielectric nanoparticles are also able to exhibit the Fano resonance~\cite{Miroshnichenko:NL:2012, Chong2014, Filonov_oligomer, Shcherbakov2015, Hopkins2013a, Hopkins:N:2013, Hopkins2015}. The important feature of such dielectric oligomers, comparing to their metallic counterparts, is the localization of the electromagnetic field inside the dielectric nanoparticles. Another important property of such oligomers is that the fundamental mode of the high-index spherical nanoparticle is magnetic dipole mode~\cite{KuznetsovSciRep2012, EvlyukhinNL2012}. Formation of this magnetic mode, as was mentioned above, is due to excitation of a circular displacement current. It occurs when the diameter of the particle is comparable to the wavelength inside the nanoparticle.

\begin{figure}[t]\centering
\includegraphics[width=0.95\columnwidth]{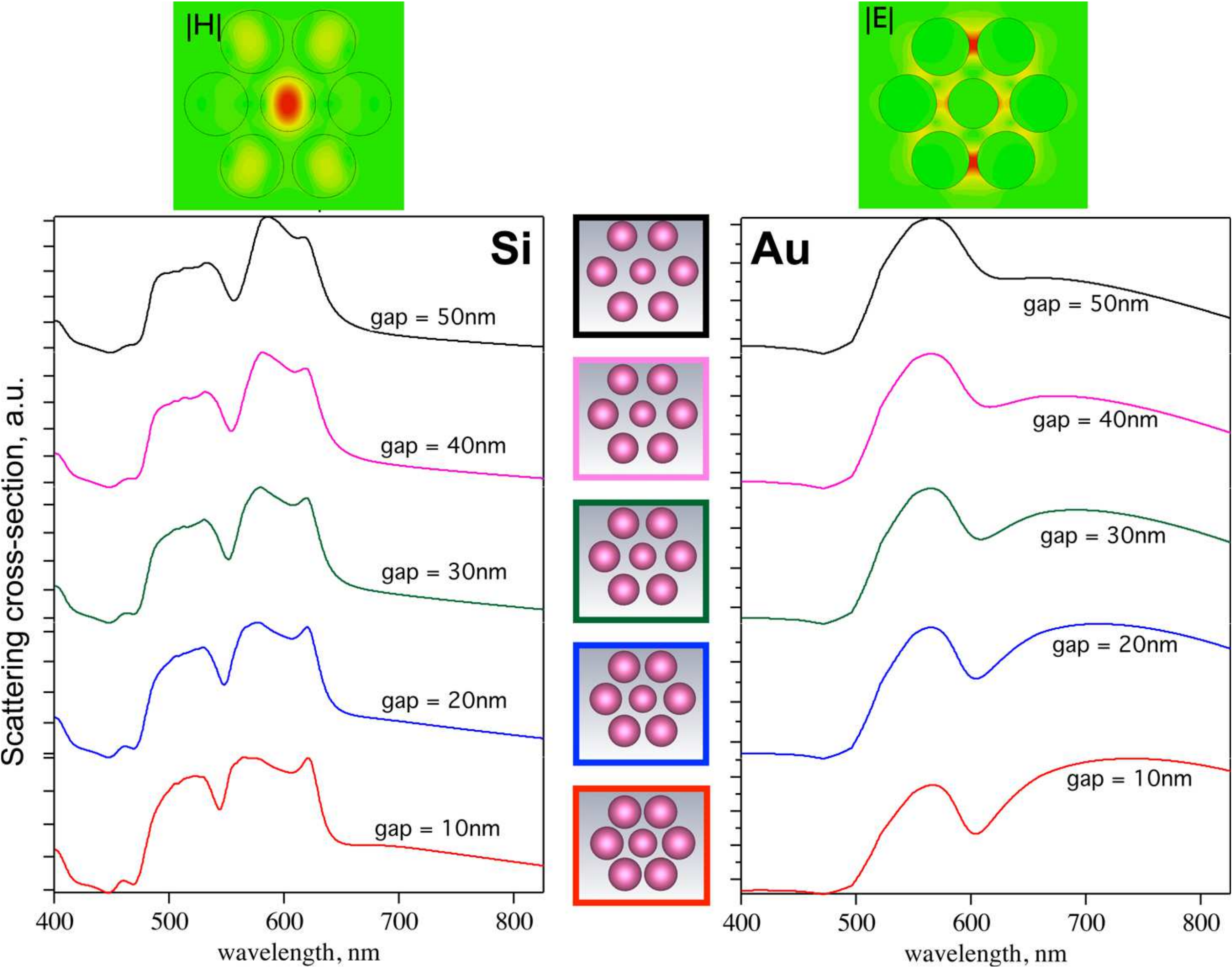}
\caption{Scattering cross-section of oligomer of silicon (left) and gold (right) nanoparticles for various separation between the particles. The radius of the central particle is R$_1$ = 65~nm and outer particles R$_2$ = 75~nm. The Fano resonance depends weakly on the separation between particles for all-dielectric oligomers, while for plasmonic analogue this dependence is very strong. It demonstrates a difference in the coupling mechanism in both situations. From Refs.~\cite{Miroshnichenko:NL:2012}.}\label{OLIgomers}
\end{figure}

Authors of Ref.~\cite{Miroshnichenko:NL:2012} have shown that the structure arranged of six identical dielectric nanoparticles, positioned in the vertices of the regular hexagon, and the particle of another radius in the center [see Figure~\ref{OLIgomers}], exhibits the Fano resonance at the resonance frequency of the central particle, while six other particles are not resonant at this frequency and they form a non-resonant mode of the whole structure. The near-field interference of this two modes leads to the suppression of the whole structure scattering and formation of the Fano resonance~\cite{Miroshnichenko:NL:2012}. In Figure~\ref{OLIgomers} the scattering cross-section spectra of oligomer of silicon (left) and gold (right) nanoparticles for various separation between the particles are presented. It has been demonstrated that the Fano resonance depends weakly on the separations between particles for all-dielectric oligomers, while for plasmonic analogue this dependence is very strong (see Figure~\ref{OLIgomers}).

\begin{figure*}[t]\centering
\includegraphics[width=0.95\textwidth]{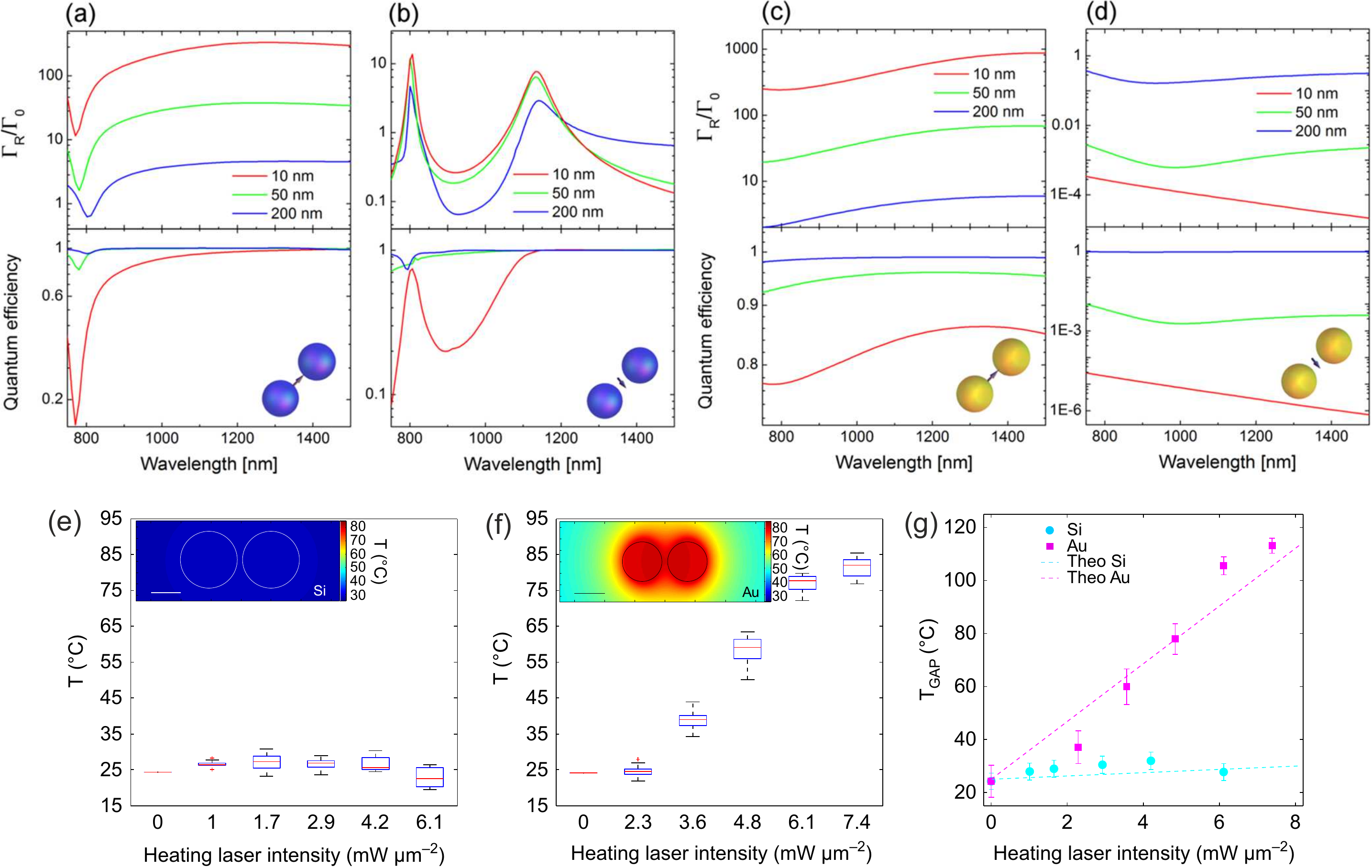}
\caption{Enhancement of the radiative decay rate and quantum efficiency of an electric dipolar emitter positioned in between two silicon ((a),(b)) and gold ((c),(d)) nanospheres of 150 nm radius. Orientations of the emitters positioned at the centers of the systems are shown in the schematics. Gap widths are given in the legends. (e), (f), (g) -- Temperature measurement in nanoantennas. Box plot for the average temperature T, measured for (e) silicon and (f) gold nanoantennas, excited at resonance. The inset in each figure shows the calculated temperature map around the disks for the heating laser intensity of 5~mW $\mu$m$^{-2}$ in both cases. Scale bar is 100 nm. (g) Extracted temperature in the gap for selected silicon (cyan) and gold (magenta) nanoantennas as a function of the heating laser intensity at 860~nm. The dashed lines show the numerical calculations for the temperature at the gap, presenting good agreement with the experimental data. The error bars show the s.d. of the temperature measurements, obtained from error propagation from the fluorescence measurements. From Refs.~\cite{Albella2013, Maier_2015}.}\label{comparison}
\end{figure*}

In Ref.~\cite{Filonov_oligomer} the existence of the Fano resonances in dielectric oligomers has been demonstrated for the first time. Due to the scalability of Maxwell equations, the authors used microwave ceramic spheres with sizes of several centimeters (instead of Si nanoparticles). Such particles exhibit the magnetic response in the microwave frequency range. The authors measured the near magnetic field in the vicinity of the dielectric oligomer with high accuracy, which allowed to verify the origin of the Fano resonance, predicted in the theoretical study~\cite{Miroshnichenko:NL:2012}.

\subsection{Surface enhanced spectroscopy and sensing}
\label{sec:sensing}

\begin{figure*}[t]\centering
\includegraphics[width=0.95\textwidth]{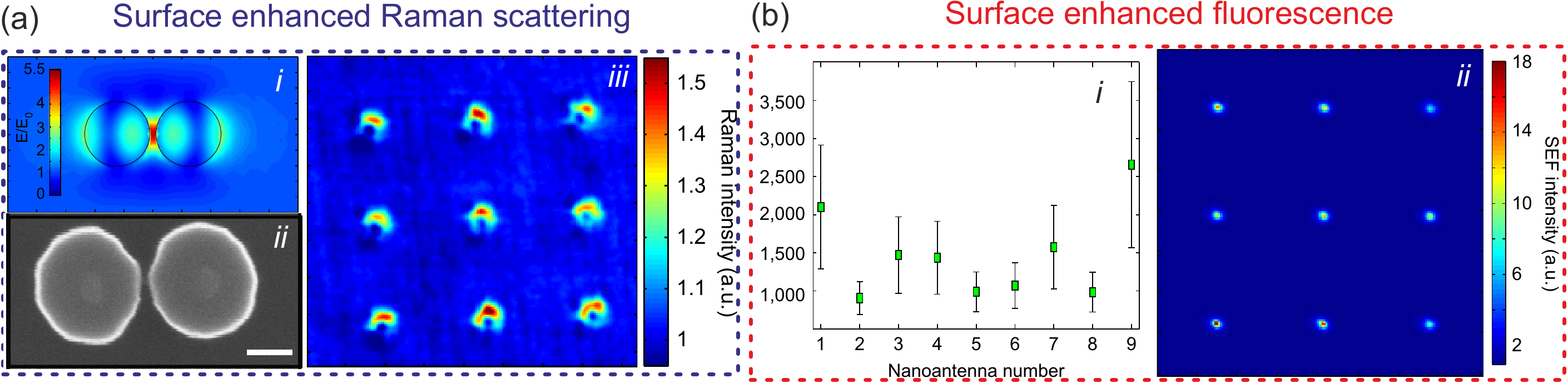}
\caption{(a)~Surface enhanced Raman scattering (SERS) via all-dielectric dimer nanoantenna. (i)~Near-field distribution map for the silicon structure excited at resonance, showing good confinement of the electric field in the gap; the maximum enhancement value is 5.5. (ii)~The SEM image of the Si nanoantenna fabricated on a silicon-on-insulator substrate. The nanoantenna consists of two identical disks with a diameter of 220~nm, a height of 200~nm, and a 20~nm gap in between. (iii)~Experimental 2D normalized Raman map, showing enhanced signal coming from the molecules close to the nanoantennas. The SERS enhancement factor reached in experiments is about 1500. (b)~Surface enhanced fluorescence via all-dielectric dimer nanoantenna. (ii)~Experimental SEF map obtained for the Si antennas. It can be clearly seen that fluorescence is enhanced over the antennas. (i)~SEF enhancement factor (FSEF) obtained from the maximum values over each antenna in (ii). The error bars show half the difference between the minimum and the maximum value in each nanoantenna. From Ref.~\cite{Maier_2015}.}\label{SERS_diel}
\end{figure*}

Resonant nanoparticles and nanostructures are proven to be powerful tools for sensing applications due to their ability to enhance and localize the optical energy in the near-field, whereas the positions of resonances depend on ambient media properties~\cite{Lal2007, Deng2015, Lei2012, Saha2012}. Moreover, nanoscale resonators provid so-called Purcell effect~\cite{Slobozhanyuk_2014APL, Krasnok_SR2015, KrasnokAPL2016, Baranov_2016_MP}, when the power radiated by a quantum light source (atom, molecula, quanum dot, etc.) is enhanced due to the increase of the local density of states (LDOS). There are numerous sensing techniques based on all these effects: surface enhanced Raman scattering (SERS)~\cite{Emory_1997, Maier_book, RuiTan2014}, surface enhanced fluorescence (SEF)~\cite{Giannini2011}, Forster resonance energy transfer~\cite{Forster_1948, DeTorres2016, Krainer2015}, refractive index sensing~\cite{Shen2013, Yang2015, Abajo_2016prl, Mesch_2016nl}, and thermometry~\cite{Lukin_2013}. Despite plasmonics has demonstrated tremendous success in the sensing applications, the concept of all-dielectric nanophotonics can also serve as a platform for high-effective sensors. First of all, low Ohmic losses in all-dielectric resonant nanostructures prevent parasitic heating of the analyzed objects~\cite{Albella2013, Albella2014, Maier_2015}, and second, high radiative part of Purcell factor and directivity improve signal extraction~\cite{Krasnok_LPR_2015}.

Figure~\ref{comparison} shows the enhancement of the radiative decay rate and quantum efficiency of an electric dipolar emitter positioned between two silicon (a--b) and gold (c--d) nanospheres, and the results of temperature measurement in such nanoantennas (e--g). Figure~\ref{comparison} (a--d) shows that the all-dielectric dimer nanoantennas have an ability to strongly enhance both electric and magnetic LDOS, whereas the plasmonic nanoantennas work only with electric dipole sources. It can be seen that the all-dielectric nanoantennas have the quantum efficiency exceeding that of the plasmonic nanoantennas. Moreover, it also can be seen that the Au nanoantennas significantly increase their temperature when the heating laser intensity increases while the Si temperature remains nearly constant and does not affect the molecules under study.

From the perspective of enhancement by individual nanoparticles, the absence of cut-off frequency for dipole plasmon resonance results in much higher field enhancement near plasmonic nanoparticles as compared to dielectric ones with same sizes in the sub-100-nm range. However, larger Si nanoparticles, possessing a magnetic Mie-type resonance at the optical frequencies, yield comparable or even larger near-field enhancement~\cite{Albella2013, Maier_2015}. This effect was proved in SERS experiments, where Si resonant nanoparticles produce larger SERS effect as compared to gold ones of the same sizes~\cite{Meseguer_2014}.

In order to get a huge local field enhancement, plasmonic dimers~\cite{Punj2015} or oligomers are used~\cite{Rahmani2013, Hentschel2010, Deng2015, Verre2015}. The same approach is also possible for dielectrics, when local field enhancement factor in the very gap of a Si dimer could be more than one order of magnitude~\cite{Bakker2015}. Such enhancement was applied to achieve high SERS and SEF effects~\cite{Albella2013, Albella2014, Maier_2015}. Therefore, all-dielectric nanostructures also provide the field enhancement, which is high enough for detection of small amount of organic materials.

Figure~\ref{SERS_diel} shows the experimental results of all-dielectric nanoantennas application for surface enhanced Raman scattering (a) and for surface enhanced fluorescence (b). It has been demonstrated that the Si-dimer nanoantennas exhibit high near-field enhancement within a 20~nm gap at the near IR wavelengths. This all-dielectric nanoantenna is able to enhance the Raman scattering of a polymer thin film by a factor of 10$^3$ (a) and also allow surface enhanced fluorescence by a factor of 2$\times$10$^3$, avoiding the well-known fluorescence quenching effects observed for metallic structures when no spacer layers are used. Moreover, the molecular thermometry measurements have demonstrated that the dielectric nanoantennas produce ultra-low heating when illuminating at their resonance wavelength, thus overcoming one of the main drawbacks of traditional plasmonic materials such as gold.

It is worth noticing that placing a detected nanoobject in a "hot spot" of a resonant nanostructure could be less effective for sensing than positioning it in a place with the highest Purcell factor. Moreover, high Purcell factor can be achieved in "cold spots" of the nanostructure~\cite{KrasnokAPL2016, Krasnok_ax_2015}. So, the conceptually different approach to extract more signal from quantum emitters is to enhance Purcell factor within the nanostructure. In case of all-dielectric nanostructures, a periodical chain of resonant nanoparticles, supporting magnetic dipole resonance, looks a promising device to achieve huge radiative part of Purcell factor for an electric dipole source~\cite{KrasnokAPL2016, Krasnok_ax_2015}, whereas dielectric material does not lead to any quenching effects.

Another important feature of the dielectric nanoparticles is the ability to shift the incident light frequency via the \textit{Raman scattering} process. The Raman scattering is inherent for dielectric materials, but almost does not exist in metals. Moreover, an intensity of the Raman scattering strongly depends on the resonant properties of dielectric nanoparticle. In particular, magnetic types of low-order Mie resonances provide much larger enhancement factors as compared to corresponding electric types of Mie resonances~\cite{Baranov16}. The enhancement factor dependence on excitation wavelength demonstrates sharp peak at a magnetic dipole resonance, demonstrating more than two orders of magnitude variation in its vicinity and narrower width ($\sim$10~nm) as compared the width of the Mie resonance ($\sim$30~nm). This effect could also be promising for a number of applications related to sensitivity of Raman signal to thermal and refractive index variations of ambient medium.

\subsection{Coupled-resonator optical waveguides}
\label{sec:waveguides}

\begin{figure*}[!t]\centering
\includegraphics[width=0.8\textwidth]{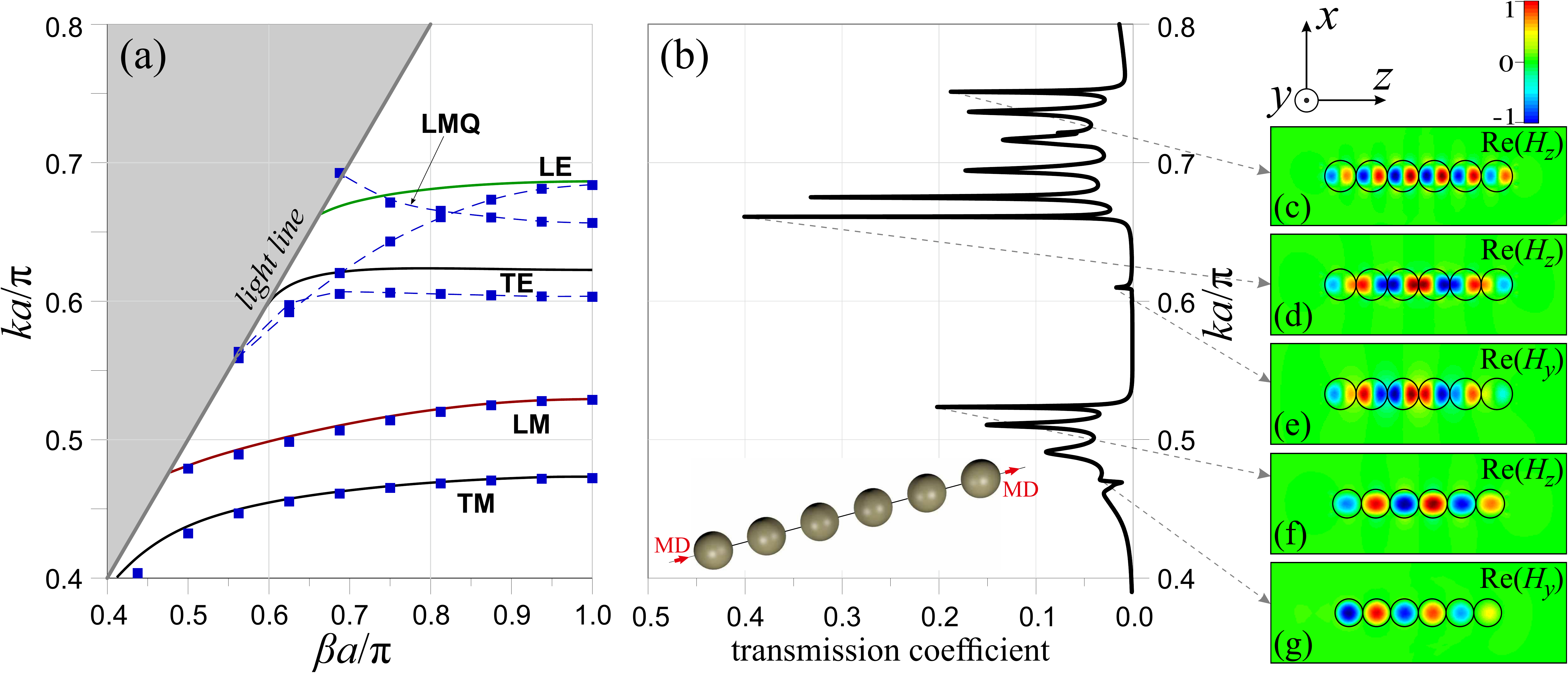}
\caption{(a)~Dispersion diagram of an infinite chain of lossless spherical Si nanoparticles with radius 70\;nm and period $a=140$\;nm. Only waveguide modes under the light line $\beta=k$ are shown. Numerically calculated eigenmodes are shown with blue squares (and thin dashed blue lines). Oblique grey line is the light line. (b)~Numerically calculated transmission spectrum of a chain of 6 Si spheres. (c--g)~Electric field distributions in the corresponding modes. The operational range of normalized frequenices $ka/\pi$ lies within an optical spectral range for the chosen parameters. Adopted from the Ref.~\cite{SavelevPRB2014}.}
\label{fig:chains_sp}
\end{figure*}

A design of highly efficient integrated circuits with combined optical and electronic components for the subwavelength guiding of the electromagnetic energy is one of the main trends of the optical communications technologies in the beginning of the 21th century~\cite{ChenSciRep2012}. In order to achieve high integration densities, optical waveguides with subwavelength light localization have been proposed. Among those are conventional Si (or other dielectric/semiconductor) nanowires, photonics crystal waveguides and plasmonic waveguides. Silicon nanowire waveguides have very small cross-section sizes, and they can be manufactured being of a high quality~\cite{LawScience2004}. However, such waveguides do not provide low-loss propagation of optical signals through sharp bends and require rather large bending geometries thus increasing the overall size of an optical chip~\cite{EspinolaOE2001}. Photonic crystals have been viewed as a possible alternative, and it has been already demonstrated that light can be guided by a waveguide composed of defects, and that such waveguides can have sharp bends~\cite{MiroshnichenkoOE2005}. However, due to the different mechanism of waveguiding, namely Bragg reflection, the overall transverse size of the photonic crystal waveguide is usually about several wavelengths. Besides, the nice property of photonic crystals to propagate light through sharp bends was found to depend strongly on the bend geometry being also linked to the strict resonant conditions associated with the Fano resonance where the waveguide bend plays a role of a specific localized defect~\cite{MiroshnichenkoOE2005}, thus demonstrating narrowband functionalities. Coventional plasmonic waveguides allow for truly subwavelength localization of light, but it is always accompanied with severe Ohmic losses, which makes the propagation lengths of surface plasmons impractically short~\cite{KhurginNN2015}.

Another candidate for the efficient subwavelength guiding is a coupled-resonator optical waveguide, where guided modes are formed by coupled resonances of the single elements~\cite{YarivOL1999}. The most recent realization of such type of waveguide was suggested in Ref.~\cite{JunjiePRA2009} in the form of the chain of high-index low-loss dielectric nanoparticles. Such waveguide was fabricated and its properties were measured several years later~\cite{BakkerProc2015}. In Ref.~\cite{SavelevPRB2014} it was shown that guiding of the electromagnetic energy in dielectric discrete waveguides is achieved due to coupled Mie resonances. And since dielectric nanoparticles support both MD and ED resonances simultaneously, waveguides composed of such nanoparticles support several modes of different types. For the case of spherical particles, first several modes (at low frequencies) calculated within the framework of the dipole approximation and via full-wave numerical simulations are shown in Figure~\ref{fig:chains_sp}(a). Two longitudinally (transversely) polarized modes marked LE and LM (TE and TM) are formed by coupling between MD and ED dipoles, oriented along (perpendicular to) the axis of the chain. Note that in the case of transverse modes there is a coupling between EDs and MDs induced in different particles, while longitudinal modes are independent of each other. However in spherical particles ED and MD resonances are well separated in frequency and, consequently, the coupling between EDs and MDs is quite small. Therefore coupled resonances of nanospheres form quasi-independent separate pass bands in different spectral ranges, due to the different dipole-dipole interaction strength.

In Figure~\ref{fig:chains_sp}(b) the results of modelling of transmission of light through the chain of 6 particles are presented. One can observe a transmission band around $ka/\pi=0.5$ formed by excited TM and LM modes [Figures~\ref{fig:chains_sp}(f,g)]. Transmission band around $ka/\pi=0.7$ is formed by multipole modes. The most high-frequency peak corresponds to the longitudinal magnetic quadrupole mode with $\beta=0$ [Figure~\ref{fig:chains_sp}(c)]. This mode crosses the light line (i.e. it is a radiating leaky wave), and therefore it is not shown in Figure~\ref{fig:chains_sp}(a), where only unattenuated modes are present. Numerically found frequency for $\beta=0$ is $ka/\pi \approx 0.76$, which coincide with the value in transmission spectrum at the upper edge of the longitudinal magnetic quadrupole band. One can also see a transmission peak at $ka/\pi \approx 0.61$ [Figure~\ref{fig:chains_sp}(e)] corresponding to the TE mode, which is also excited due to the inhomogenity of current in the probes.

\begin{figure}[t]\centering
\includegraphics[width=0.99\columnwidth]{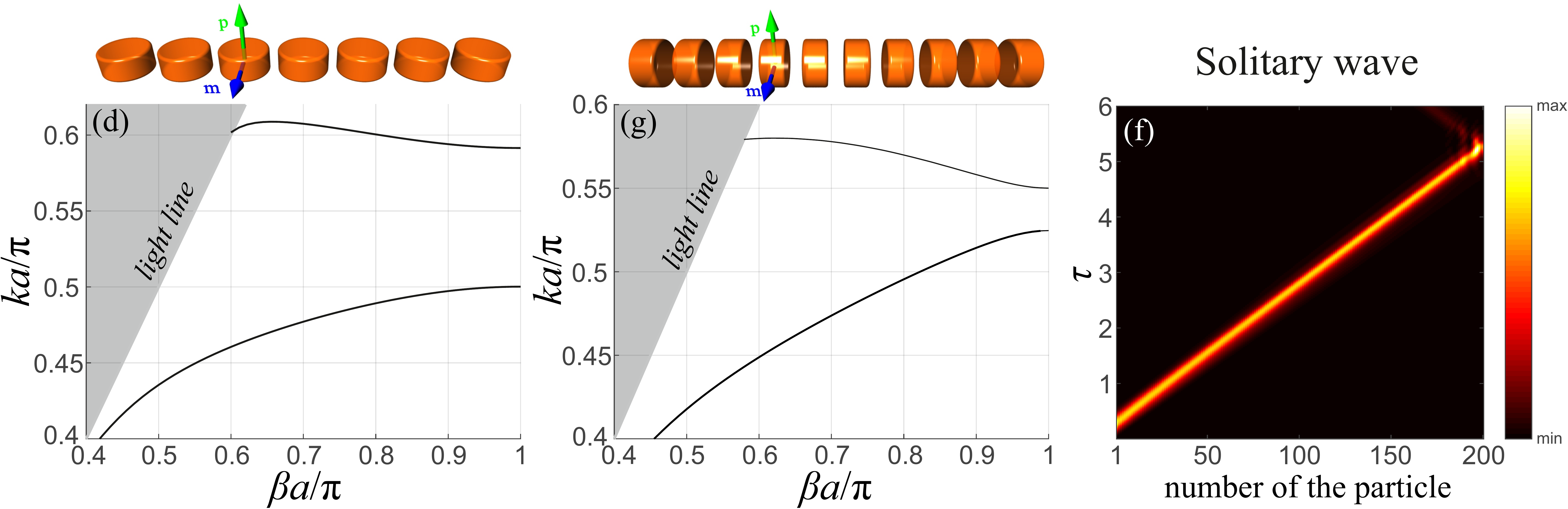}
\caption{(a,b) Numerically calculated dispersion curves for the transversely polarized modes of an infinite chain of dielectric nanocylinders (a) with radius $R=90$\;nm and height $h=180$\;nm and (b) with radius $R=130$\;nm and height $h=120$\;nm; period is $a=200$\;nm in all cases. $\beta a/\pi$ is the normalized Bloch wave number and $ka/\pi$ is the normalized frequency. (c) Calculated nonlinear dynamics of the pulse propagating in the chain of dielectric nanocylinders with dispersion properties shown in (b) in the nonlinear regime, excited with a Gaussian 100 fs pulse with normalized center frequency 0.553. Adopted from the Ref.~\cite{Savelev_ACSP_2016}.}
\label{fig:chains_nonsp}
\end{figure}

For nonspherical particles, resonance frequencies depend on the orientation of the dipole oscillations, and the corresponding passbands can be shifted by changing the particle parameters. Besides the shifting of the operational frequency range, change of the resonance frequencies can also substantially increase the coupling between EDs and MDs of the neighboring particles. This happens for transversely polarized modes when resonance frequencies of EDs and MDs oriented perpendicular both to the axis of the chain and to each other get closer. In Figures~\ref{fig:chains_nonsp}(a,b) it is shown that second branch of the chain of cylindrical nanoparticles with close ED and MD resonance frequencies changes the sign of group velocity and group velocity dispersion in a certain frequency range. One can see, that while it is not possible for the chain of spherical nanoparticles, the interplay of the ED and MD resonances in the cylindrical nanoparticles with certain parameters can induce the anomalous dispersion regime in the discrete waveguide. This feature was employed in the study of the nonlinear regimes of femtosecond optical pulse propagation through all-dielectric waveguides~\cite{Savelev_ACSP_2016}. It has been shown that for the chains of cylindrical particles the broadening of the propagating pulse can be compensated by the nonlinear Kerr effect, thus making possible the formation and propagation of \textit{solitary waves}.

In Ref.~\cite{Savelev2014_1} the transmission efficiency of the discrete waveguides composed of arrays of high-index dielectric nanodisks with and without sharp bends has been studied. The appropriate period of the chain has been chosen so that longitudinal and transverse pass bands of the nanodisk chain, formed by coupled MD resonances of nanodisks, overlap (which cannot be done with spherical particles). This condition allows to realize an efficient transmission through sharp 90$^\circ$ bends. In the straight chain of nanodisks two passbands formed by Fabry-Perot resonances of mixed LM-TM modes and TE modes of the finite chain were observed in transmission efficiency spectrum. In the waveguide composed of 30 disks with a 90$^\circ$ bend in the middle the maximum value of the transmission efficiency of about 0.6 was achieved. Also it has been shown that the LM mode in the horizontal branch can transform into the TM mode propagating in the vertical branch and vice versa. These theoretical conclusions have been supported by presenting the experimental results for the microwave frequencies for an efficient guiding through 90$^\circ$ bend in a microwave dielectric waveguide.

\begin{figure}[!t]\centering
\includegraphics[width=0.99\columnwidth]{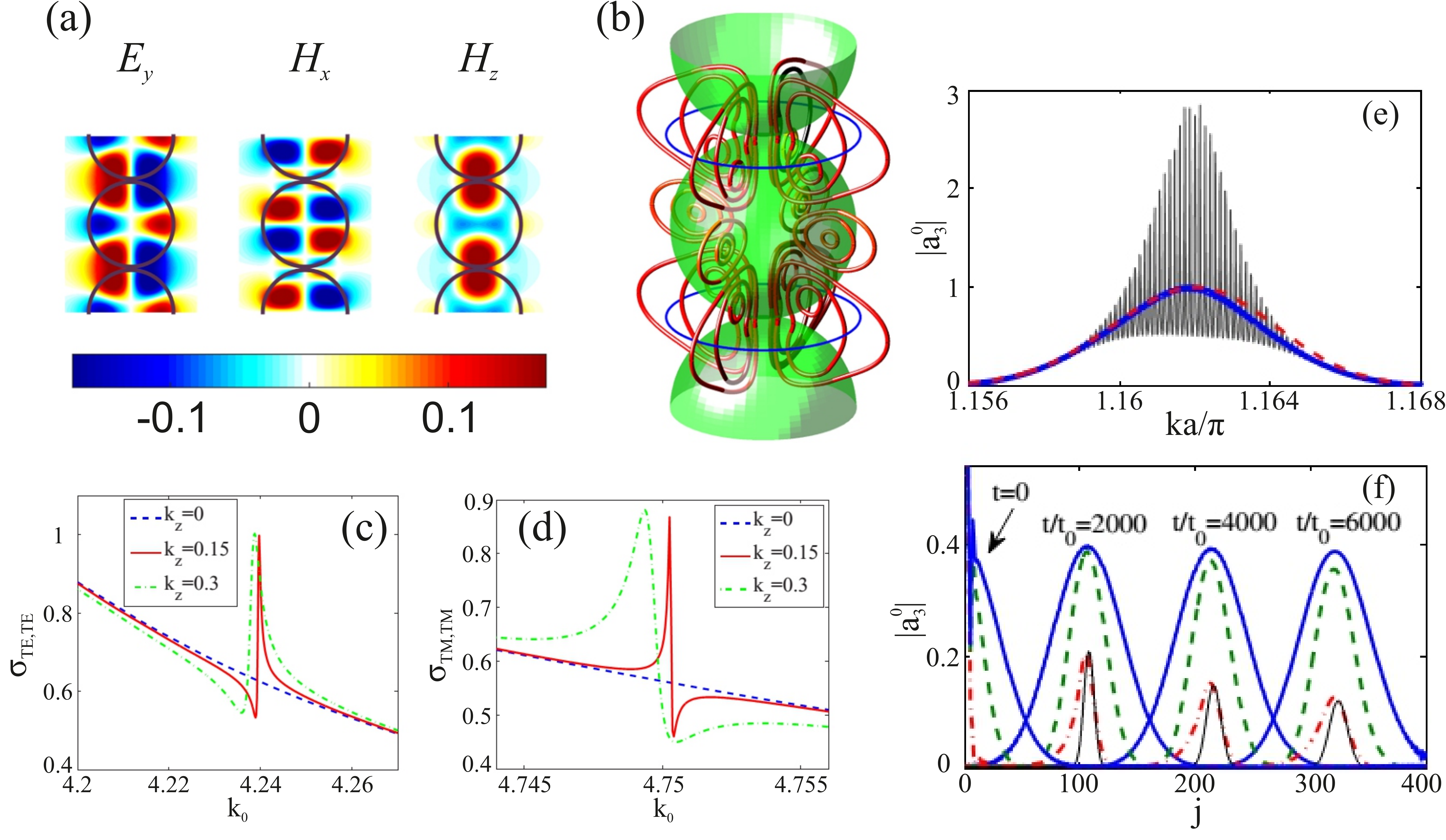}
\caption{(a) The real parts of the EM field components and (b) the electric force lines (red) and
the magnetic force lines (blue) of the symmetry-protected bound state embedded into the TM continuum with zero azimuthal number and zero Bloch wavenumber. (c,d) Total scattering cross section of the array illuminated by a plane wave. Scattering of the (c) TE and (d) TM plane waves is strongly affected by the presence of the symmetry-protected BSC. (e,f) Light propagation in the array of 400 nanoparticles. Absolute value of the leading Mie coefficient versus (e) wave number of the stationary wave injected into the array and (f) the particle number $j$ for a light pulse with different pulse widths. Adopted from the Refs.~\cite{BulgakovPRA2015,BulgakovOL2016}.}
\label{fig:chains_BICs}
\end{figure}

Recently, several studies suggested the chains of high-index dielectric nanoparticles as a simple 1D platform for studying \textit{bound states in the continuum} (BICs)~\cite{BulgakovPRA2015, BulgakovOL2016}. The BICs are the localized states that exist within the continuum spectrum of radiative waves~\cite{Shabanov, BIC}. In the Ref.~\cite{BulgakovPRA2015} TE and TM BICs in a linear periodic array of dielectric spheres were demonstrated. Field distribution and force lines of the simplest TM BIC in a zero diffraction channel with zero azimuthal wavenumber and zero Bloch wavenumber are shown in Figures~\ref{fig:chains_BICs}(a,b), respectively. Different TE- and TM-polarized BICs with non-zero azimuthal numbers and non-zero Bloch wavenumbers were also predicted. Such peculiar states manifest themselves in the spectra of scattering cross-section of a plane wave by infinite array of nanoparticles. For the parameters of the system close to the existence of a BIC, Fano-type resonance occurs due to the interference between the plane wave and the localized mode of the array [see dashed green and solid red curves in Figures~\ref{fig:chains_BICs}(c,d)]. For the parameters exactly corresponding to the BIC state, quality factor of the localized state tends to infinity which results in the collapse of the Fano resonance [see dashed blue curves in Figures~\ref{fig:chains_BICs}(c,d)].

In the Ref.~\cite{BulgakovOL2016} light guiding above the light line in the chain of dielectric nanospheres was also demonstrated. It was shown that at the frequencies close to the BICs light can propagate to the large distances in both stationary and pulse regimes. In the stationary regime the possibility of waveguiding at different frequencies can be determined from the standing waves formed in finite chain of 400 nanoparticles [Figure~\ref{fig:chains_BICs}(e)]. In Figure~\ref{fig:chains_BICs}(f) moments induced in nanoparticles during the pulse propagation are shown for different pulse widths. Such calculations revealed that pulses with a certain width and the central frequency tuned exactly to the frequency of a BIC can propagate in the chains of dielectric nanospheres for the distances of tens and hundreds wavelengths.

Thus, chains of dielectric nanoparticles provides one with the simple and efficient platform, allowing guiding and localization of light in linear and nonlinear regimes. The discrete waveguides based on high-index dielectric nanoparticles may exceed its currently existing analogs: plasmonic waveguides, dielectric photonic crystals, and homogeneous Si waveguides, offering a large number of customizable options and negligible energy dissipation. Such waveguides can be used in photonic components responsible for the transmission of information in the optical and optoelectronic integrated circuits.

\subsection{Metamaterials and metasurfaces}
\label{sec:metamaterials}

Future technologies will push for a steep increase in photonic integration and energy efficiency, far surpassing that of bulk optical components, Si photonics, and plasmonic circuits. Such level of integration can be achieved by embedding the data processing and waveguiding functionalities at the level of material rather than a chip, and the only possible solution to meet those challenges is to employ the recently-emerged concept of {\it metamaterials and metasurfaces}. Metamaterials are artificial media with exotic electromagnetic properties not available in natural media which are specially created in order to reach functionalities required for particular applications~\cite{Engheta2014, KivsharNM2012}. Metasurfaces are their two-dimensional implementations that are much simpler for fabrication~\cite{Shalaev_2011}. Metamaterials have been studied since 2000 and revealed such effects as negative refraction, backward waves, beating of diffraction limit (subwavelength imaging), and became a paradigm for engineering electromagnetic space and controlling propagation of waves by means of transformation optics~\cite{Engheta2014, Shalaev_2011, ZayatsNP2012}. The research agenda is now focusing on achieving tunable, switchable, nonlinear and sensing functionalities of metamaterials. Since 2010 the studies have been shifted to the stage of practical implementation and development of real metadevices. As a result, a novel concept of metadevices, that can be defined as metamaterial-based devices with novel and useful functionalities achieved by the structuring of functional matter on the subwavelength scale, has been developed.~\cite{KivsharNM2012}. The metadevices practical implementation is the general trend in the area of metamaterials.

\begin{figure}[!t] \centering
\includegraphics[width=0.99\columnwidth]{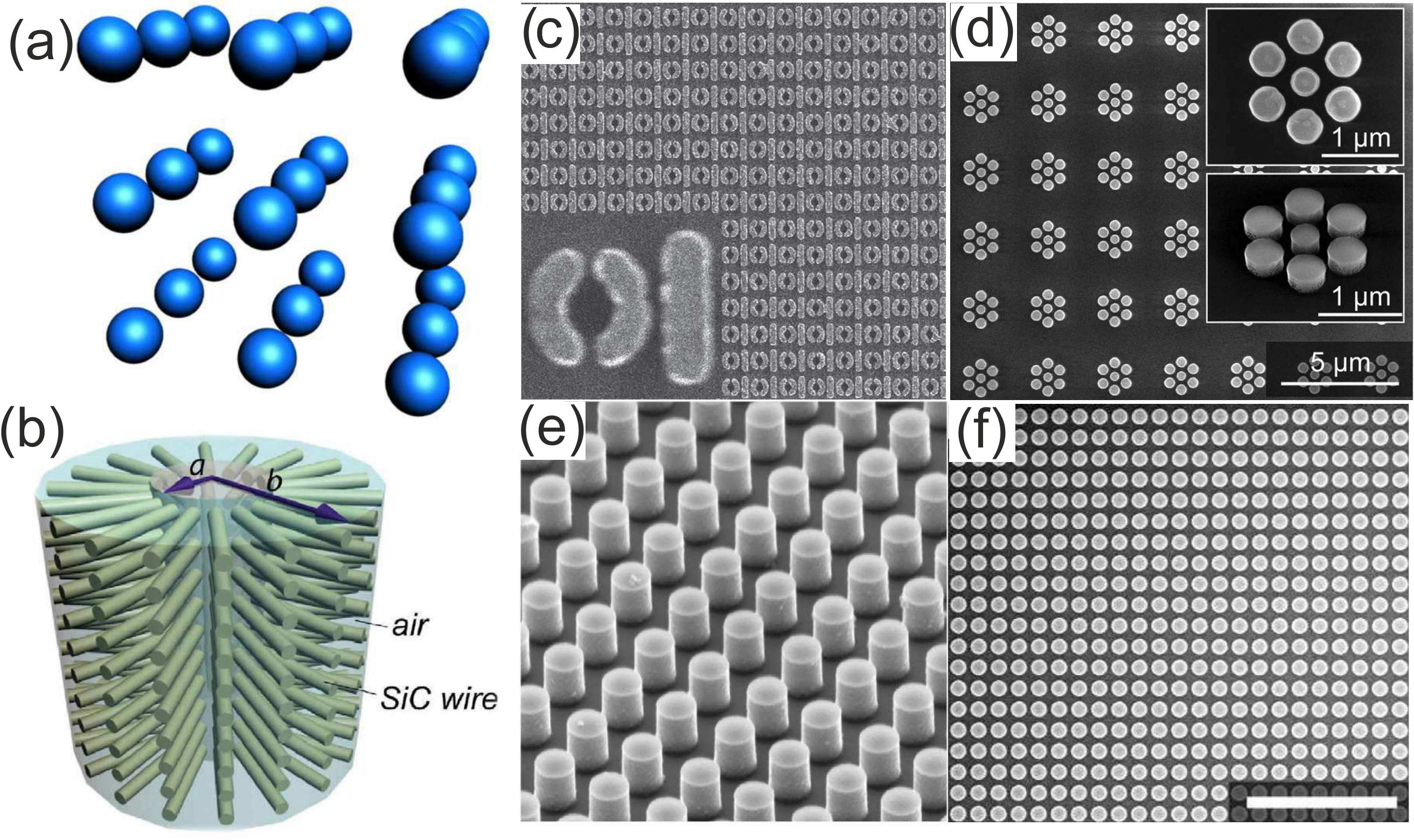}
\caption{(a,b) All-dielectric metamaterials based on spherical and cylindrical particles~\cite{Mosallaei2008, Zhao09, Odit2009, TretyakovJAP2006}; (c-f) all-dielectric metasurfaces~\cite{Staude_15, SpinelliNC2012, Valentine2014}.}
\label{fig:meta}
\end{figure}

The area of metamaterials has opened a broad range of technologically important capabilities ranging from the subwavelength focusing to ``stopped light'', including their ability to control magnetic response of novel subwavelength structured materials. This is important because the magnetic response of natural materials at optical frequencies is very weak due to the diminishing of electronic spin states at high frequencies~\cite{Landau_CM}. That is why only the electric component of light is directly controlled in conventional photonic devices. However, effective control of light at the nanoscale requires the presence of the electric and magnetic responses, simultaneously. A vast majority of the current metamaterial structures exhibiting magnetic response contain metallic elements with high conductive losses at optical frequencies, which limits their performance. As it was mentioned above, one of the canonical examples is a split-ring resonator that is an inductive metallic ring with a gap that is a building block of many metamaterials. This concept works very well for gigahertz~\cite{Smith, Shelby}, terahertz~\cite{Padilla06} and even near-infrared (few hundreds~THz)~\cite{Liu08} frequencies. However, for shorter wavelengths and in particular for the visible spectral range, this concept fails due to the dissipative losses and fabrication difficulties~\cite{Soukoulis07}.

In order to overcome these fundamental problems, an alternative approach of all-dielectric metamaterials has been proposed~\cite{Peng2007, ShullerPRL2007, Meng2008, Mosallaei2008, Guizal2009, Zhao09, Cummer_08, Brener_12, Lukyanchuk13, Zhou2014, Valentine2014, Limonov2015, Jahani2016}. In this case, a high-index dielectric particle, exhibiting magnetic and electric Mie resonances plays a role of a single \textit{meta-atom}. Such high-index dielectric particles replace their metallic counterparts in metamaterials and metasurfaces. For example, it has been shown that the 3D dielectric composite of high-index dielectric particles [as shown in Figure~\ref{fig:meta}(a)] exhibits the negative permeability near the first Mie resonance~\cite{Meng2008, Mosallaei2008}. Even more complex geometry of such all-dielectric metamaterials have been proposed in Refs.~\cite{Odit2009, TretyakovJAP2006}. For theoretical treatment of such composites consisting of high-index dielectric spherical particles embedded in a low-index dielectric matrix the Levin's model can be used~\cite{Lewin_1947, Mosallaei2008}.

Conventional optical components rely on gradual phase shifts accumulated during light propagation to shape light beams. The nanostructured design can introduce new degrees of freedom by making abrupt phase changes over the scale of the wavelength. A two-dimensional lattice of optical resonators or nanoantennas on a planar surface, with the spatially varying phase response and subwavelength separation, can imprint such phase patterns and discontinuities on propagating light as it traverses the interface between two media. In this regime, anomalous reflection and refraction phenomena can be observed in optically thin metamaterial layers, or optical metasurfaces (see Figure~\ref{fig:meta}(c--f)), creating surfaces with unique functionalities and engineered reflection and transmission laws. The first example of such metasurface -- a lattice of metallic nanoantennas on Si with a linear phase variation along the interface, was demonstrated recently~\cite{Gaburro11, Zhang12NL, Tsai12NL, AluPRL2013, Grbic2013}. The concept of metasurfaces with phase discontinuities allows introducing generalized laws for light reflection, and such surfaces provide great flexibility in the control of light beams, being also associated with the generation of optical vortices. Metasurfaces can also be used for the implementation of important applications such as light bending ~\cite{Gaburro11} and specific lenses~\cite{Capasso12}.

The phase gradient metasurfaces created by high-index nanoparticles of varying shape have been recently proposed~\cite{Staude_15, Capasso_2015}. In the article~\cite{Staude_15}, for the first time, highly efficient all-dielectric metasurfaces for near-infrared frequencies using arrays of Si nanodisks as meta-atoms have been proposed and realized [see Figure~\ref{fig:meta}(f)]. The authors have employed the main features of Huygens' sources, namely spectrally overlapping electric and magnetic dipole resonances of equal strength, to demonstrate Huygens' metasurfaces with a full transmission-phase coverage of 360 degrees and near-unity transmission, and confirmed experimentally full phase coverage combined with high efficiency in transmission. Based on these key properties, the authors show that all-dielectric Huygens' metasurfaces could become a new paradigm for flat optical devices, including beam-steering, beam-shaping, and focusing, as well as holography and dispersion control.

We also should note here that the studies in the field of dielectric metamaterials in microwave frequency range performed until 2009 are summarized in the review paper~\cite{Zhao2009}. The current state of research in this area given in Refs.~\cite{Jacob_optica_2014, Corbitt2014, Krasnok_SPIE2015, Savelev2015} and especially in the excellent Review paper by S.~Jahani and Zubin Jacob~\cite{Jahani2016}.

\subsection{Nonlinear nanophotonics applications}
\label{sec:nonlinear}

An enhancement of nonlinear optical response at the nanoscale is also challenging area of nanooptics, where dielectric materials have already been implemented for various micro-devices. In particular, Raman lasing~\cite{Vahala_2002, Freude2010}, supercontinuum generation~\cite{Freude2010}, and all-optical switching~\cite{Freude2010, ZayatsNP2012, Noskov2012} are the bright examples of nonlinear photonics based on Si micro-devices (waveguides, ring-resonators, photonic crystals etc.). Indeed, inherent nonlinear response of many dielectrics (especially, semiconductors) is very high in the optical and IR range, being comparable with metals or even much larger due to non-centrosymmetrical crystalline lattice of some dielectric materials (GaAs, GaP, Te, etc.). On the other hand, plasmonic nanodevices paved the way to creation of deeply subwavelength nonlinear devices~\cite{ZayatsNP2012, Krasavin2016}. Therefore, implementation of plasmonic principles for developing of all-dielectric nonlinear nanodevices looks tempting. Recently, the enhancement of optical nonlinearities in resonant Si nanostructures has been demonstrated theoretically and experimentally at the scale of single nanoparticles~\cite{Noskov2012, ShcherbakovNL2014, Shcherbakov2015NL, Makarov2015, Valentine2015, Iyer2015, Baranov_2016_ACS, Baranov_dimer_2016}.

\begin{figure}[t]\centering
\includegraphics[width=0.99\columnwidth]{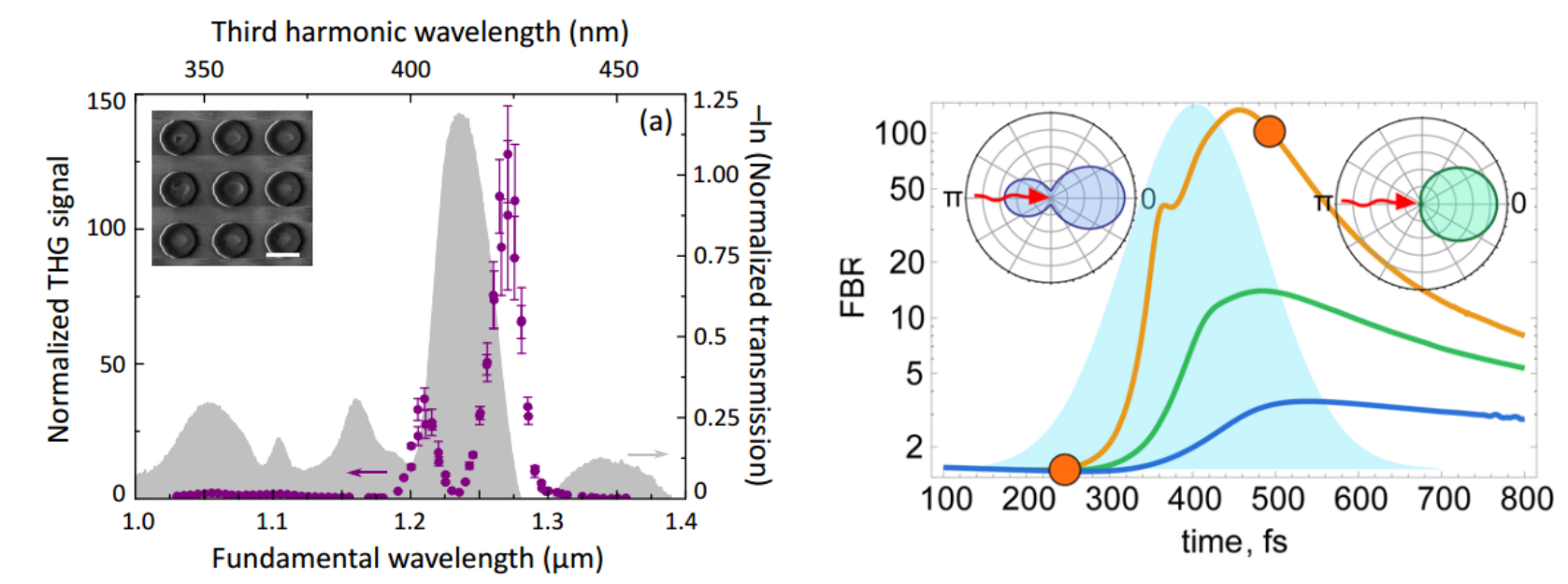}
\caption{(a)~Purple dots: third harmonic generation spectrum of the nanodisks array (shown in the inset). The gray area represents the linear transmission spectrum. (b)~Dynamical reconfiguration of a Si nanoantenna directivity via photoexcitation: Front-to-Back ratio of a nanoparticle during the action of a 200 fs pulse with peak intensities of 10 GW/cm$^2$ (blue), 20 GW/cm$^2$ (green) and 40 GW/cm$^2$ (orange). Scattering diagrams of the incident beam at the largest intensity are shown in the two insets.}
\label{nonlinear}
\end{figure}

One of the most attractive applications of all-dielectric nanostructures is the efficient \textit{frequency conversion}. In the pioneering work on this topic~\cite{ShcherbakovNL2014} enhancement of third-harmonic generation from Si nanoparticles (in form of nanodisks) exhibiting both electric and magnetic dipolar resonances has been demonstrated, Figure~\ref{nonlinear}(a). The efficiency of IR-to-visible conversion by 2 orders of magnitude in the vicinity of the magnetic dipole resonance with respect to the unstructured bulk Si slab was achieved. The idea of the conversion enhancement at the magnetic resonance has been developed in subsequent works with regard to the generation of higher optical harmonics~\cite{Valentine2015, Shorokhov_2016, Shcherbakov2015, Smirnova_ACS_2016} and Raman scattering~\cite{Baranov16}. Dielectric oligomers~\cite{Shorokhov_2016} and nanoparticles supporting the anapole mode excitation~\cite{Anapole} have also been employed for third harmonic generation enhancement.

Another intriguing non-linear effect arising in resonant Si nanoparticles is the electron-hole plasma photoexcitation. Silicon is a semiconductor, and at normal conditions, its conduction band is almost empty. However, optical absorption causes the electrons to fill the conduction band thus altering its permittivity and optical response~\cite{Sokolowski-Tinten2000}. Recently, the photoexcitation of plasma was employed for tuning of Si nanoantenna optical properties in the IR and visible regions~\cite{Makarov2015, Valentine2015, Baranov_2016_ACS, Baranov_dimer_2016}. It has been shown that the plasma photoexcitation allows for manipulating of electric and magnetic nanoparticle responses, resulting in dramatic changes of both scattering diagram and scattering cross section, Figure~\ref{nonlinear}(b). The 20\% tuning of reflectance of a single Si nanoparticle by femtosecond laser pulses with the wavelength in the vicinity of the magnetic dipole resonance has been demonstrated. In the recent work~\cite{Baranov_dimer_2016} this effect has been utilized for achieving a pronounced \textit{beam steering effect} in an all-dielectric dimer nanoantenna.

Last but not least, resonant dielectric nanoparticles demonstrate \textit{much higher damage threshold}. For comparison, the typical values of damage threshold of metallic nanostructures are: gold nanorods ($\sim$70~GW/cm$^2$ or $\sim$10~mJ/cm$^2$ at 130~fs~\cite{Zayats_NN2011}), gold G-shaped nanostructures ($\sim$ 100~GW/cm$^2$ or $\sim$3~mJ/cm$^2$ at 30~fs~\cite{Valev2012}), and gold nanocylinders ($\sim$200~GW/cm$^2$ or $\sim$20~mJ/cm$^2$ at 100~fs~\cite{Zuev_2016}). According to the known data from the literature, low-loss Si nanoparticles have significantly higher damage threshold: $\sim$400~GW/cm$^2$ or $\sim$100~mJ/cm$^2$ at 250~fs~\cite{Valentine2015}; and $\sim$1000~GW/cm$^2$ or $\sim$100~mJ/cm$^2$ at 100~fs~\cite{Makarov2015}. Such considerable difference in damage thresholds for plasmonic (e.g. gold) and all-dielectric (e.g. Si) materials originates form difference in their melting temperatures (Tm(Au)=1337~K and Tm(Si)=1687~K), and enthalpies of fusion (H(Au)=12.5~kJ/mol and H(Si)=50.2~kJ/mol). Therefore, silicon-based nanostructures are more stable than plasmonic ones upon intense laser irradiation, which makes them very attractive for nonlinear applications.\\

\section{Conclusions and outlook}

In this Article, we have reviewed some of the recent developments in the field of all-dielectric nanophotonics. This area of optical science studies the light interaction with high-index dielectric nanoparticles supporting optically-induced electric and magnetic Mie resonances. We have described several advances in this field which demonstrate that dielectric structures allow to control both magnetic and electric components of light in a desirable way, and also discuss properties of high-indexed nanoparticles along with their fabrication methods. We have reviewed the practical applications area of all-dielectric nanophotonics, including the nanoantennas for the quantum source emission engineering, the all-dielectric oligomers and their Fano resonances, the surface enhanced spectroscopy and sensing, coupled-resonator optical waveguides, optical solitons and bound-states-in-continuum, all-dielectric metamaterials and metasurfaces, and the nonlinear nanophotonics.

Future technologies will demand a huge increase in photonic integration and energy efficiency far surpassing that of bulk optical components and silicon photonics. Such an integration can be achieved by embedding the data-processing and waveguiding functionalities at the material level, creating the new paradigm of metadevices. It is now believed that robust and reliable metadevices will allow photonics to compete with electronics not only in telecommunication systems but also at the level of consumer products. The main challenges in achieving this goal will be in developing cost-efficient fabrication and device integration technologies. All-dielectric nanophotonics is seen as a practical way to implement many of the important concepts of nanophotonics allowing high functionalities and low-loss performance of metadevices.

\section*{acknowledgement}
	This work was financially supported by project GZ~3.4424.2017/HM.


%

\end{document}